\documentclass{aa}  
\usepackage{graphicx}
\usepackage{txfonts}
\usepackage{xcolor}
\usepackage{hyperref}
\usepackage[normalem]{ulem}
\defcitealias{2026arXiv260500093J}{J26}
\hypersetup{colorlinks=true,urlcolor=blue,linkcolor=blue,citecolor=blue}

\def \MSUN{\rm M_{\odot}}

\def \sics{\sigma_{\rm T}/\rm m_{\chi}}
\newcommand{\mytilde}{\raise.19ex\hbox{$\scriptstyle\sim$}}

\begin{document} 

   \title{TNGC-SIDM I: Evolution of massive cluster mergers under the self-interacting dark matter paradigm}
   \titlerunning{TNGC-SIDM I: SIDM probes in merging clusters}

   \author{W. Lee \inst{1,2}\thanks{\email{wonkilee@mpia.de}}, 
   D. Nelson \inst{3,4},
   M. J. Jee \inst{2,5}, 
   G. Despali \inst{6,7,8},
   M. Vogelsberger \inst{9,10},
   A. Pillepich \inst{1},
   C. Roche \inst{9},
   K. Finner \inst{11},
   H. Joo \inst{2},
   D. Nagai \inst{12},   
   J. ZuHone \inst{13}
   }
    
   \authorrunning{W. Lee et al.}

    \institute{Max-Planck-Institut f{\"u}r Astronomie, K{\"o}nigstuhl 17, D-69117 Heidelberg, Germany 
    \and 
    Yonsei University, Department of Astronomy, Seoul, Republic of Korea 
    \and
    Zentrum f{\"u}r Astronomie der Universit{\"a}t Heidelberg, ITA, Albert Ueberle Str. 2, D-69120 Heidelberg, Germany
    \and
    Universit{\"a}t Heidelberg, Interdisziplin{\"a}res Zentrum f{\"u}r Wissenschaftliches Rechnen, INF 205, 69120 Heidelberg, Germany
    \and 
    Department of Physics, University of California, Davis, One Shields Avenue, Davis, CA 95616, USA
    \and 
    Dipartimento di Fisica e Astronomia “Augusto Righi”, Alma Mater Studiorum Università di Bologna, Via Gobetti 93/2, I-40129 Bologna, Italy
    \and 
    INAF-Osservatorio di Astrofisica e Scienza dello Spazio di Bologna, Via Piero Gobetti 93/3, I-40129 Bologna, Italy
    \and 
    INFN-Sezione di Bologna, Viale Berti Pichat 6/2, I-40127 Bologna, Italy
    \and 
    Department of Physics and Kavli Institute for Astrophysics and Space Research, Massachusetts Institute of Technology, Cambridge, MA 02139, USA
    \and 
    Fachbereich Physik, Philipps Universit\"at Marburg, D-35032 Marburg, Germany
    \and 
    IPAC, California Institute of Technology, 1200 E California Blvd., Pasadena, CA 91125, USA
    \and 
    Department of Physics, Yale University, New Haven, CT 06520, USA
    \and 
    Harvard-Smithsonian Center for Astrophysics, 60 Garden St., Cambridge, MA 02138, USA
    }
   %
    %
    %
    %
             
  \date{}

  \abstract{
  
  Merging galaxy clusters provide an ideal laboratory for testing the self-interacting nature of dark matter. In this paper, we present a pilot study of magnetohydrodynamical cosmological zoom-in simulations of two cluster mergers selected from TNG-Cluster, with $M_{\rm 200c} \sim 9\times10^{14}~M_\odot$ and $M_{\rm 200c}\sim 2\times10^{14}~M_\odot$. The simulations are performed with both standard collisionless cold dark matter (CDM) and self-interacting dark matter (SIDM), assuming an isotropic, velocity-independent cross-section of $\sigma_{\rm T}/m_\chi = 1~\rm cm^2~g^{-1}$, with an additional $0.5~\rm cm^2~g^{-1}$ run for the low-mass system, and the mass resolution and galaxy formation model are unchanged from TNG-Cluster. We show how dark matter self-interactions produce an isothermal core in the dark matter halo, generate unique substructures in the intracluster light, and introduce variation in the multi-wavelength observables of the intracluster plasma. Using these simulations, we test the validity of observational probes for dark matter self-interactions in cluster mergers. We find that the shock-to-shock separation shows similar evolution in the self-interaction models tested here, providing a reference for distinguishing the $\mytilde100-200~\rm kpc$-scale difference in halo separation induced by self-interactions. Moreover, the dark matter halo centroid of the main cluster in the massive cluster merger is offset from the brightest cluster galaxy by as much as $\mytilde20~\rm kpc$, larger than the offset in the sub-cluster. This offset persists beyond 4 Gyr after the first pericenter passage. Our results indicate that sample selection plays a key role in testing the possible self-interacting nature of dark matter using cluster mergers, and that double radio relic cluster mergers and massive clusters with evidence of an ancient major merger offer a promising testbed for this purpose.
       }

   \maketitle
%

\section{Introduction}
\label{sec:intro}

Dark matter plays a fundamental role in cosmic structure formation, providing the gravitational potential within which galaxies and galaxy clusters assemble over cosmic time. The collisionless cold dark matter (CDM) model has been broadly successful in reproducing the observed large-scale structure of the universe \citep[e.g.,][]{2005Natur.435..629S,2016A&A...594A..13P}. However, on small scales, several discrepancies between CDM predictions and observations have been reported \citep[][]{2017ARA&A..55..343B}, including the core-cusp problem \citep[e.g.,][]{2001ApJ...552L..23D}, the diversity of dwarf galaxy rotation curves \citep[e.g.,][]{2015MNRAS.452.3650O}, the too-big-to-fail problems \citep[e.g.,][]{2011MNRAS.415L..40B}, and an overabundance of galaxy-galaxy strong lensing in massive clusters \citep[e.g.,][]{2020Sci...369.1347M,2023A&A...678L...2M}. These discrepancies may reflect the baryonic feedback processes not fully captured in these simulations \citep[e.g.,][]{2010Natur.463..203G,2013ApJ...765...22B,2022A&A...668A.188M}. Alternatively, since dark matter particle candidates have yet to be reported from direct \citep[e.g.,][]{2018PhRvL.121k1302A} or indirect \citep[e.g.,][]{2024PhRvD.109f3024M} detection experiments, this may indicate the variations in the underlying nature of dark matter \citep[e.g.,][]{2012MNRAS.420.2318L,2012MNRAS.423.3740V,2016MNRAS.460.1399V,2025ApJ...978...38D}.

Self-interacting dark matter (SIDM) is one of the leading alternative models proposed to address these small-scale discrepancies \citep{2000PhRvL..84.3760S}. Unlike CDM, SIDM allows dark matter particles to exchange momentum through non-gravitational scattering, characterized by the self-interaction cross-section per unit mass $\sigma_{\rm T}/m_{\chi}$. Because the scattering rate scales with the local dark matter density, self-interactions are most frequent in the central regions of halos while remaining negligible in low-density environments \citep[e.g.,][]{2013MNRAS.430...81R}. This makes SIDM consistent with the successful large-scale structure predictions of CDM, since the linear evolution of density perturbations is unaffected by self-interactions \citep{2016PhRvD..93l3527C}, while modifying the inner halo structure in ways that can alleviate small-scale tensions \citep[for recent reviews, see][]{2018PhR...730....1T,2025RvMP...97d5004A}.

Astrophysical probes across different mass and velocity scales are used to constrain $\sics$ \citep[e.g.,][]{2025RvMP...97d5004A}. Among these probes, galaxy clusters have been widely used to explore dark matter self-interaction in the high-velocity regime, with their $10^{14-15}~M_{\odot}$ of dark matter within a relatively small volume. Multiple observable signatures of dark matter self-interaction have been proposed at cluster scales. The density profile \citep[$\sics\lesssim0.3~\rm cm^{2}~g^{-1}$; e.g.,][]{2016PhRvL.116d1302K,2021JCAP...01..024S,2022A&A...666A..41E}, halo shapes \citep[$\sics\lesssim1~\rm cm^{2}~g^{-1}$; e.g.,][]{2002ApJ...564...60M,2013MNRAS.430..105P,2018MNRAS.474..746B}, and the spatial offset between the brightest cluster galaxy (BCG) and the cluster center \citep[$\sics\lesssim0.39~\rm cm^{2}~g^{-1}$; e.g.,][]{2017MNRAS.469.1414K,2019MNRAS.488.1572H} commonly favor a small self-interaction cross-section in galaxy clusters. Dwarf-galaxy scales, in contrast, often favor a larger cross-section that can reproduce the observed density or velocity profiles \citep[e.g.,][]{2013MNRAS.431L..20Z,2015MNRAS.453...29E,2017PhRvL.119k1102K,2017MNRAS.468.2283C,2026NatAs..10..440V}. As these systems provide environments with different DM relative velocities \citep[e.g.,][]{2024MNRAS.529.2327F}, these independent constraints from different astrophysical probes can further constrain the velocity-dependent nature of SIDM.

Merging galaxy clusters provide a unique laboratory for probing dark matter self-interaction in high-velocity environments. During a cluster merger, the collision between dark matter halos maximizes the probability of dark matter particle encounters, with a high encounter velocity due to the collision velocity of clusters $\mytilde2000-4000~\rm km~s^{-1}$ \citep[e.g.,][]{2002ApJ...567L..27M}. The boosted self-interactions during the pericenter passage exert additional drag force on dark matter halos and reduce the halo separation relative to their collisionless galaxy and stellar populations, while the same momentum exchange drives active evaporation that reduces their mass \citep[e.g.,][]{2000PhRvL..84.3760S,2014MNRAS.437.2865K}. Using this behavior, early studies constrained the $\sics$ in merging clusters by measuring the spatial offset between the BCG or galaxy distribution and the mass distribution inferred from gravitational lensing \citep[e.g.,][]{2004ApJ...606..819M,2012ApJ...747L..42D,2014ApJ...783...78J}, combined with the mass-to-light ratio \citep[M/L;][]{2008ApJ...679.1173R}, or placed an upper limit based on the survived sub-cluster core \citep[e.g.,][]{2004ApJ...606..819M}. These probes yielded a tight upper limit in the high-velocity environment, either with the textbook example of merging clusters \citep[e.g., the Bullet Cluster, $\sics\lesssim0.2\rm~cm^2~g^ {-1}$;][]{2004ApJ...606..819M,2008ApJ...679.1173R,2025ApJ...987L..15C} or from the statistical analysis of the galaxy-dark matter halo offset \citep[e.g., $\sics\lesssim0.47\rm~cm^2~g^ {-1}$;][]{2015Sci...347.1462H}. However, later studies claim that these galaxy-based probes in merging clusters are likely limited by the boundedness of galaxies to the dark matter potential \citep[e.g.,][]{2017MNRAS.469.1414K}. Moreover, the merger phase can introduce an additional bias and relax the upper limit from the offset statistics when it is accounted for \citep[$\sics\lesssim2\rm~cm^2~g^{-1}$;][]{2018ApJ...869..104W}. Since colliding clusters observed at a given separation could be caught either in their outgoing phase, when the halo separation increases with time after pericenter passage, or in their returning phase, when the separation decreases with time after apocenter passage, an accurate merger history reconstruction was needed for using merging clusters as SIDM testbeds.

Recently, \citet[][]{2026arXiv260500093J} proposed a complementary approach utilizing radio relics. Radio relics are diffuse synchrotron emissions observed at cluster outskirts, whose Mpc-scale extent, thin widths, and high polarization fractions have been attributed to a merger shock acceleration origin \citep[e.g.,][]{2019SSRv..215...16V}. \citet{2026arXiv260500093J} showed that the shock-to-shock distance traced by double radio relics remains nearly independent of the self-interaction cross-section, since the collision velocity is set by the total mass of the system and self-interaction does not alter the collision process before the shock is launched \citep[e.g.,][]{2024MNRAS.529.2327F}. Comparing the halo-to-halo separation at a fixed relic-to-relic separation therefore isolates the effect of self-interaction, since the halo separation decreases with increasing cross-section after pericenter passage. Using a Bayesian inference that compares the observed halo and relic separations in eleven double radio relic cluster mergers with a suite of idealized binary merger simulations, \citet{2026arXiv260500093J} derived an upper limit of $\sics \lesssim 0.22~\rm cm^{2}~g^{-1}$ with $68\%$ confidence, one of the tightest constraints on the self-interaction cross-section from merging clusters to date.

Numerical simulations remain the primary tool for predicting these observable signatures of self-interaction in cluster mergers. Since observations cannot provide a controlled test with varying dark matter physics, multiple numerical simulations have been performed with the same initial conditions but varying SIDM models to track their effects on the observables. Previous studies have used different simulation setups depending on the primary focus of their study. Idealized simulations analyze the impact of a specific physical parameter under simplified, isolated conditions \citep[e.g.,][]{2014MNRAS.437.2865K,2021MNRAS.505..851F,2026arXiv260500093J}, or reproduce specific observed systems given their flexibility in tuning parameters \citep[e.g.,][]{2017MNRAS.465..569R}. Cosmological simulations follow the evolution inside a full cosmological box, enabling statistical analyses under a more realistic mass accretion history and large-scale structure \citep[e.g.,][]{2013MNRAS.430...81R,2019MNRAS.488.3646R,2025A&A...697A.213D,2026A&A...708A..47D, 2026A&A...706A.340G}. Cosmological zoom-in simulations re-simulate the target of interest at high resolution, reaching a mass resolution comparable to idealized simulations while capturing the large-scale structure in low-resolution \citep[e.g.,][]{2018MNRAS.476L..20R,2024A&A...687A.270R,2025A&A...703A.112H,2025ApJ...991...69N}. These numerical simulations have also explored various self-interaction models, including frequent and rare self-interaction \citep[e.g.,][]{2014MNRAS.437.2865K,2021MNRAS.505..851F}, isotropic and anisotropic interaction \citep[e.g.,][]{2017MNRAS.467.4719R}, and velocity-independent and -dependent cross-sections \citep[e.g.,][]{2012MNRAS.423.3740V,2021MNRAS.503..920C,2024MNRAS.529.2327F,2025ApJ...991...69N}.

Despite this range of numerical works, the impact of SIDM in massive cluster mergers remains poorly explored under a realistic cosmological setup, mainly due to the rarity of these systems. From observations, the main constraints are often provided by massive cluster mergers on the $10^{15}M_{\odot}$ scale \citep[e.g.,][]{2025ApJ...987L..15C}. However, only a limited number of such massive clusters appear in cosmological boxes. Moreover, although all massive clusters experience mergers across their formation history, the collision between two massive clusters is relatively rare \citep[e.g.,][]{2010ApJ...718...60L}. Thus, simulating realistic massive cluster mergers in a full cosmological box requires a Gpc-scale volume \citep[e.g.,][]{2023MNRAS.526.4978S}, which can be computationally demanding to achieve high spatial resolution. As a result, SIDM probes of merging clusters often rely on predictions from idealized simulations \citep[e.g.,][]{2014MNRAS.437.2865K,2017MNRAS.465..569R,2017MNRAS.467.4719R,2021MNRAS.505..851F,2023MNRAS.523.5915F,2026arXiv260500093J}, which may create model bias when quantitatively comparing with observed cluster mergers.

In this regard, we present the TNG-Cluster SIDM project, hereafter TNGC-SIDM, a spin-off of TNG-Cluster \citep{2024A&A...686A.157N} that specifically re-simulates merging clusters selected from the original TNG-Cluster runs with various SIDM models. Covering a range of masses and merger parameters, this sample of cluster mergers simulated with and without SIDM aims to provide a controlled testbed for evaluating existing methods and predicting new signatures to constrain $\sics$ in merging clusters. The validated methods can then be applied to the statistical samples of multi-wavelength observations of merging clusters \citep[e.g.,][]{2019ApJS..240...39G,2025ApJS..277...28F}, which are expected to expand substantially with upcoming wide-field surveys such as the SKA and the Rubin Observatory. In this work we present a pilot study of two magnetohydrodynamical cosmological zoom-in simulations of two cluster mergers, including one massive cluster with $M_{\rm 200c}\sim9\times10^{14}~M_{\odot}$\footnote{$R_{\Delta\rm c}$ denotes the radius within which the mean density equals $\Delta$ times the critical density of the universe, and $M_{\Delta\rm c}$ is the total mass enclosed within $R_{\Delta\rm c}$.} and a low-mass cluster $M_{\rm 200c}\sim2\times10^{14}~M_{\odot}$, experiencing a major merger with a mass ratio of $\mytilde1:3$ and $\mytilde1:2$, respectively. These clusters are re-simulated with CDM and SIDM models assuming isotropic, velocity-independent cross-sections. We examine the multi-wavelength observables of the simulated systems across dark matter models and confirm that relic-to-relic separation provides a reliable reference for testing the SIDM scenario, and the BCG and the dark halo centroids present long-lasting spatial offsets that can be detected even when the cluster relaxes after the collision.

This paper is organized as follows. In Section~\ref{sec:method}, we describe the simulations and the methods used to construct the multi-wavelength observables. Section~\ref{sec:result} presents an overview of the simulated clusters and their mergers, followed by an examination of the effectiveness of the SIDM probes proposed by previous studies in Section~\ref{sec: SIDM_tracers}. We discuss the resolution and projection effects on these SIDM probes in Section~\ref{sec: discussion}, before we summarize our findings in Section~\ref{sec: summary}. Throughout, we adopt the Planck 2015 $\Lambda$CDM cosmology \citep{2016A&A...594A..13P}, with $h=0.6774$, $\Omega_{\rm m}=0.3089$, $\Omega_{\rm \Lambda}=0.6911$, $\Omega_{\rm b}=0.0486$, $\sigma_{8}=0.8159$, and $n_{\rm s}=0.9667$. 


\section{Method}
\label{sec:method}

\subsection{Cosmological zoom-in simulations of merging cluster with SIDM}
\label{sec:tngc-sidm}

We perform cosmological zoom-in simulations of a merging galaxy cluster using the initial conditions from TNG-Cluster. TNG-Cluster is a suite of cosmological magnetohydrodynamical (MHD) zoom-in simulations of 352 massive galaxy clusters selected from a parent volume of $(1~\rm Gpc)^3$ \citep{2024A&A...686A.157N}. Our SIDM runs adopt the same initial conditions as the original TNG-Cluster runs while implementing the SIDM models. As self-interactions act in dense environments, the SIDM runs present the same cluster merger event found in the CDM runs, while the merger history after the collision is modified as self-interactions are enhanced during the pericenter passage. This allows us to use the merger catalog of TNG-Cluster \citep[][]{2024A&A...686A..55L} to select clusters ideal for testing SIDM. 

Self-interactions between the dark matter particles are implemented in \texttt{AREPO} following \citet{2012MNRAS.423.3740V}. Specifically, we adopt an elastic scattering model with the isotropic, velocity-independent cross-section, $\sics$. The scattering probability is determined by the cross-section, the local dark matter density, and the relative velocity between particles \citep[see][for details]{2012MNRAS.423.3740V}. We re-simulate the same cluster simulations with the CDM and SIDM models, with $\sigma_{\rm T}/m_{\chi}=1~\rm cm^{2}~g^{-1}$. We note that the adopted cross-sections exceed the upper limits derived from cluster-scale probes for a constant, velocity-independent model \citep[e.g.,][]{2008ApJ...679.1173R,2015Sci...347.1462H,2025ApJ...987L..15C,2026arXiv260500093J}, or the effective cross-section expected from velocity-dependent SIDM models at cluster collision velocities \citep[e.g.,][]{2021MNRAS.503..920C,2024MNRAS.529.2327F}. Nevertheless, we adopt these cross-sections to demonstrate the influence of self-interaction on multi-wavelength observables and to test the observational tracers proposed by earlier studies. 

Following observational strategies of selecting cluster mergers for SIDM constraints \citep[e.g.,][]{2026arXiv260500093J}, we re-simulate clusters that experienced a binary, major cluster merger with no third cluster- or group-scale halos that can modify the signatures. In this work, we select two cluster mergers: a massive, high-velocity merger and a low-mass merger. The massive cluster has $M_{\rm 200c}=8.8\times10^{14}~M_{\odot}$ ($M_{\rm 500c}=6.6\times10^{14}~M_{\odot}$) at $z=0$ and experiences a major merger at $z=0.39$ with a mass ratio of $\mytilde1:3$. The cluster separation at their first pericenter passage is $\mytilde80~\rm kpc$ ($\mytilde0.04R_{\rm 200c}$) with the collision velocity\footnote{The collision velocity is derived using the relative average velocity of the stellar particles within 100~kpc of each halo center. The velocity is comparable when we define this using dark matter particles.} of $\mytilde\rm 3000~km~s^{-1}$, making this system a Bullet-like head-on high-velocity collider. The cluster does not experience an additional merger event after the major merger, and so the cluster has accumulated its mass comparable to that at $z=0$ already by $z=0.35$. As a result, this system further allows us to track the long-term evolution of merging clusters in the SIDM paradigm without further confusion with another significant merger. The massive cluster has also experienced a minor merger with a mass ratio of $\sim0.05$ at $z=0.7$ with a group-scale halo ($M_{\rm 200c}=2\times10^{13}~M_{\odot}$). Both major and minor mergers produce Mpc-size merger shocks, but only the major merger presents double radio relics, while the minor merger generates a single radio relic in front of the sub-cluster. We re-simulate this cluster with two DM models: CDM ($\tt M9S0$) and SIDM with $\sics=1~\rm cm^{2}~g^{-1}$ ($\tt M9S1$). 

For the low-mass cluster merger, the system has a mass of $M_{\rm 200c}=2\times10^{14}~M_{\odot}$ ($M_{\rm 500c}=1.5\times10^{14}~M_{\odot}$) at $z=0$ and experiences its first pericenter passage at $z=0.14$ with a mass ratio of $\mytilde0.5$. The first pericenter separation is $\mytilde130~\rm kpc$ ($\mytilde0.1R_{\rm 200c}$), and the collision velocity between the clusters is $\mytilde\rm 1400~km~s^{-1}$. Similar to the massive cluster merger, the low-mass cluster merger also produces double radio relics after the collision, with the two relics showing comparable luminosity and size to each other. We note that the clusters have reached their second pericenter passage at $z=0$, and so we can only explore the first outgoing and returning phases. 
For this cluster, we perform simulations with three different DM models: CDM ($\tt M2S0$), SIDM with $\sigma_{\rm T}/m_{\chi}=0.5~\rm cm^{2}~g^{-1}$ ($\tt M2S05$), and SIDM with $\sigma_{\rm T}/m_{\chi}=1~\rm cm^{2}~g^{-1}$ ($\tt M2S1$).  

The numerical resolution and the subgrid physics follow that of TNG-Cluster. The simulations are performed with the moving-mesh code \texttt{AREPO} \citep{2010MNRAS.401..791S}, including the intracluster magnetic field evolution \citep{2011MNRAS.418.1392P, 2013MNRAS.432..176P} and the galaxy formation model implemented following the IllustrisTNG framework \citep{2017MNRAS.465.3291W, 2018MNRAS.473.4077P}. We utilize the shock properties from the on-the-fly shock finder for modeling the synchrotron radio emission generated by shock acceleration \citep{2015MNRAS.446.3992S, 2016MNRAS.461.4441S,2024A&A...686A..55L}. Dark matter and baryons are resolved with particle masses of \mbox{$m_{\rm DM}\sim6\times10^{7}~\MSUN$} and \mbox{$m_{\rm baryon}\sim10^{7}~\MSUN$}, respectively. The gravitational softening length for dark matter and stars is $\mytilde1.4~\rm kpc$ at $z=0$, while gas cells use an adaptive comoving softening with a minimum physical scale of $\mytilde0.4~\rm kpc$, allowing us to accurately track the merger shocks in the cluster outskirts \citep{2024A&A...686A..55L}. We confirm that no low-resolution dark matter particles outside the zoom-in regions have fallen within $2R_{\rm200c}$ of the cluster center in all simulations. 


In addition to the main runs, we also perform low-resolution runs with and without SIDM implementation, with the resolution consistent with that of TNG300-2 \citep{2019ComAC...6....2N}, to test the dependence on the numerical resolution. We also run the dark-matter-only (DMO) counterparts, which are compared with the main runs to distinguish the role of baryons from that of self-interactions. Table~\ref{tab:simulations} summarizes the simulation runs analyzed in this paper. Hereafter, we refer to each simulation by the name listed in this table. 

\begin{table}
    \caption{Summary of the performed simulations.}
    \centering
    \scriptsize
    \begin{tabular}{cccccccc}
           \hline
           \hline
           Name &  Type & $\sigma_{\rm T}/m_{\chi}$ & $m_{\rm DM}$ & $m_{\rm baryon}$ & $M_{\rm 200c,~merger}$ & $z_{\rm merger}$ \\  
            & & $[\rm cm^2~g^{-1}]$ & $[10^7~M_{\odot}]$ & $[10^7~M_{\odot}]$ & $[10^{14}~M_{\odot}]$ & \\
           \hline
           \texttt{M9S1}    & MHD &  1.0 & 6.1 & 1.1 & 8.48 & 0.39 \\ 
           \texttt{M9S0}    & MHD &  0.0 & 6.1 & 1.1 & 8.49 & 0.39\\ 
           \texttt{M2S1}    & MHD &  1.0 & 6.1 & 1.1 & 2.1 & 0.14 \\ 
           \texttt{M2S05}   & MHD &  0.5 & 6.1 & 1.1 & 2.1 & 0.14 \\ 
           \texttt{M2S0}    & MHD &  0.0 & 6.1 & 1.1 & 2.1 & 0.14  \\ 
           \texttt{M9S1Low} & MHD &  1.0 & 48. & 8.9 & 8.5 & 0.39 \\ 
           \texttt{M9S0Low} & MHD &  0.0 & 48. & 8.9 & 8.5 & 0.39 \\ 
           \texttt{M2S1Low} & MHD &  1.0 & 48. & 8.9 & 2.1 & 0.14 \\ 
           \texttt{M2S0Low} & MHD &  0.0 & 48. & 8.9 & 2.1 & 0.14 \\ 
           \hline
           
           \texttt{M9S1-D} & DMO & 1.0 & 7.2 & 0 & 8.44 & 0.39  \\ 
           \texttt{M9S0-D} & DMO & 0.0 & 7.2 & 0 & 8.42 & 0.39  \\ 
           \texttt{M2S1-D} & DMO & 1.0 & 7.2 & 0 & 2.1 & 0.14 \\ 
           \texttt{M2S0-D} & DMO & 0.0 & 7.2 & 0 & 2.1 & 0.14 \\ 
           
            \hline
    \end{tabular}
    \tablefoot{Column 1: Simulation name. Column 2: Simulation type with DMO: dark matter-only simulations, MHD: magnetohydrodynamical simulations with galaxy formation models. Column 3: Self-interaction cross-section. Column 4: Dark matter particle mass. Column 5: Baryon particle mass. Column 6: Main cluster mass at the first pericenter passage. Column 7: Redshift of the first pericenter passage. 
    } 
    \label{tab:simulations}
\end{table}

\subsection{Multi-wavelength observables of merging clusters}
\label{sec: mwfeature}

In this work, we aim to provide an overview of these simulated SIDM galaxy clusters and test observational tracers of SIDM proposed for merging clusters. We examine three tracers that have been discussed in the previous studies: relic-to-relic separation \citep[][]{2026arXiv260500093J}, the BCG-dark matter halo offset \citep[e.g.,][]{2004ApJ...606..819M,2008ApJ...679.1173R, 2014MNRAS.437.2865K,2015Sci...347.1462H,2017MNRAS.469.1414K,2024OJAp....7E..65R}, and the mass-to-light ratio \citep[e.g.,][]{2008ApJ...679.1173R,2017MNRAS.469.1414K}. 

When comparing the observable properties of merging clusters, we forward-model the simulated clusters to obtain observable properties from the projection maps. Throughout this manuscript, we project the systems such that the system is viewed as a plane-of-sky merger, in which merging clusters are preferentially selected \citep[e.g.,][]{2019ApJS..240...39G}. For the massive cluster merger, we project the system such that the relative velocity vector of the merging clusters in the \texttt{M9S0} run aligns with the x-axis of the projection maps. The same projection axis has been applied to all the \texttt{M9} runs, as the collision axis is consistent between the runs. The low-mass cluster merger occurs along the y-axis of the simulation box, and so we project the \texttt{M2} runs along its z-axis.

We measure the relic-to-relic separation from projected radio emission maps generated from the simulated merger shocks. Following \citet{2024A&A...686A..55L,2026ApJ...998..201L}, cells associated with individual radio relics are identified based on their spatial connectivity in three dimensions. Observations estimate the relic separation using the outer edge of the diffuse radio emissions to track the shock front \citep[e.g.,][]{2022ApJ...924...18L}. To mimic this, for each identified relic, we generate a projected radio surface brightness map and construct the cumulative surface brightness profile with decreasing clustercentric distance. The profile is built along the axis connecting the cluster center of mass to the radio emissivity-weighted relic center, with an opening angle of 20$^\circ$. The distance to the relic front is defined as the radius at which the cumulative surface brightness reaches 10\% of the total, following \citet{2026ApJ...998..201L}, which corresponds to the outermost edge of the radio emission where the surface brightness drops. We confirm that the location traces the relic edge with visual inspection and verify that the relic separation is insensitive to different choices of opening angle, with consistent separation presented for opening angles of 10$^\circ$ and 30$^\circ$.

The relics propagate outward, and their luminosity decreases with time after the pericenter passage \citep[e.g.,][]{2018ApJ...857...26H,2024A&A...690A.146N}. As we identify simulated double relics based on their luminosity and spatial extent, faint diffuse emission unassociated with the merger could be misclassified as double relics at the late merger phase. To avoid this confusion, we track relic positions only while their luminosity satisfies $L_{\rm 1.4~GHz}>10^{22}~\rm W~Hz^{-1}$ and $>10^{21}~\rm W~Hz^{-1}$ for the massive and low-mass cluster mergers, respectively. We note that these lower limits are an order of magnitude below the relic luminosity predicted by the observed \citep[][]{2014MNRAS.444.3130D} and simulated \citep[][]{2024A&A...686A..55L} mass-luminosity relations, and fainter than the faintest radio relic reported in a similar mass range \citep[e.g.,][]{2022ApJ...925...91S}.

The spatial offset between the dark matter halo center and the BCG is measured from projected dark matter and stellar mass maps. Dark matter particles are smoothed using Gaussian kernels with widths set to half the distance enclosing 64 neighboring particles, while stellar particles are smoothed using Gaussian kernels with widths set to the distance enclosing 32 neighboring particles. For ICM-related observables, including X-ray and radio surface brightness, gas cells are smoothed using a Gaussian kernel of width $(m_i/\rho_i)^{1/3}$. The BCG in each subcluster is identified as the subhalo with the largest stellar mass within twice its stellar half-mass radius. As the system undergoes a binary cluster merger, we track two BCGs throughout the merger, each associated with one of the merging subclusters.

We also generate projection maps using particles that are bound to the cluster but not to the subhalos. The particle members of the cluster are defined using the \texttt{SUBFIND} \citep{2001MNRAS.328..726S} algorithm, which assigns membership based on gravitational boundedness to the local potential. We apply this selection when the dark matter and stellar mass contents of an interloping subhalo can bias our measurements, for example, when defining the center of mass to determine the BCG-dark matter halo offset. As this selects particles bound to the cluster potential but not to individual galaxies, the selected stellar population can be interpreted as the combination of the BCG and the intracluster light \citep[ICL, e.g.,][]{2019MNRAS.482.2838M}. The ICL defined with this selection, however, may include stellar subhalos with disrupted dark matter halos, especially in the SIDM runs. As this would typically be masked in observations \citep[e.g.,][]{2023Natur.613...37J}, our selection using \texttt{SUBFIND} may yield a broader ICL definition than that commonly adopted in observational studies. An observation-like definition of the ICL and its properties will be discussed in the follow-up paper in more detail.

\begin{figure}
 \centering
 \includegraphics[width=\columnwidth]{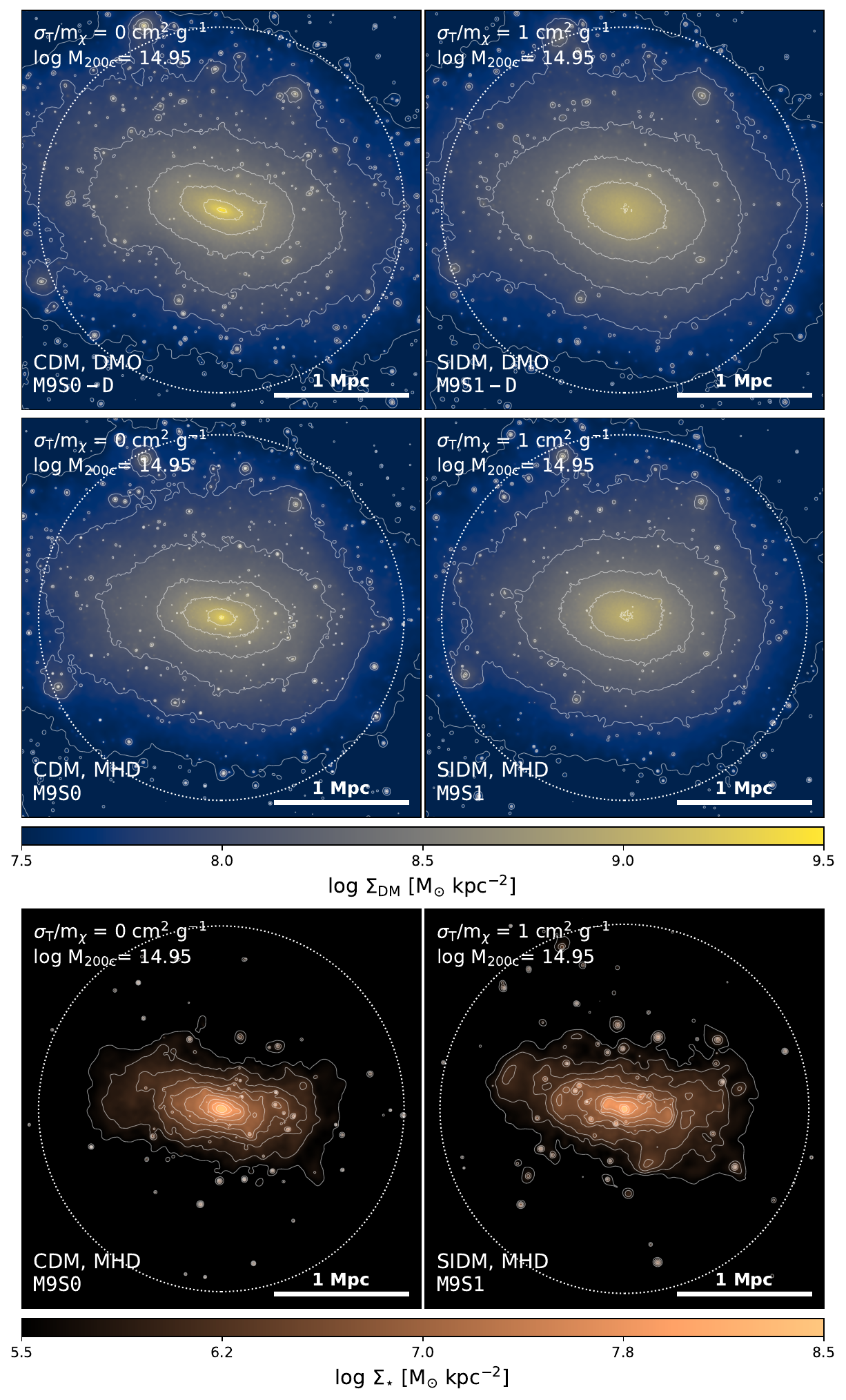}
 \caption{
 Dark matter surface density in the dark-matter-only (top) and MHD simulations (middle), and stellar surface density distribution in the MHD simulations (bottom row) of the massive galaxy cluster runs at $z=0$. The stellar density represents the combined populations of BCG and ICL, selected based on \texttt{SUBFIND}. The left and right columns show the collisionless and self-interacting dark matter runs with $\sics=1~\rm cm^{2}~g^{-1}$, respectively. Contours are spaced by 0.3 on a log scale in both the dark matter and stellar mass distributions, with the same levels adopted for the CDM and SIDM runs. The annotated circle marks $R_{\rm 500c}$ of the cluster. The snapshot presents $\mytilde4.36\rm~Gyr$ after the first pericenter passage of the major merger event, whose collision axis is aligned with the x-axis of the image. The self-interactions smooth the central dark matter density and reduce the number of subhalos, while the overall stellar mass distribution remains consistent.
 }
 \label{fig:DMmap_SIDM}
\end{figure}
\section{Galaxy clusters and mergers in the CDM and SIDM paradigms}
\label{sec:result}

The CDM and SIDM simulations present the multi-wavelength morphology of the massive and low-mass cluster mergers under different dark matter models. As the massive cluster does not experience an additional merger after the major merger at $z=0.39$, the system further allows us to examine the SIDM features in a relatively relaxed state. In this section, we describe the overall features of the CDM and SIDM clusters and their evolution during the mergers.

\subsection{Dark matter and stellar mass distributions}
\label{sec:SIDM_halo}

The top panel of Figure~\ref{fig:DMmap_SIDM} presents the dark matter distribution in the DMO simulations, \texttt{M9S0-D} and \texttt{M9S1-D}, and in the MHD simulations, \texttt{M9S0} and \texttt{M9S1}, at $z=0.0$. The presented snapshots show $\mytilde4.36\rm~Gyr$ after the first pericenter passage of the major cluster merger. At this epoch, the cluster merger process has been completed in all simulations and presents a single cluster with $M_{\rm 200c}\sim8.8\times10^{14}~M_{\odot}$. We note that, although two clusters have merged in all runs, some multi-wavelength features discussed hereafter may arise from the different merger completion times. Specifically, the sub-cluster completely merges with the main cluster at $z\sim0.03$ (time since collision (TSC)$\sim4.0\rm~Gyr$) and $ z\sim0.15$ ($\rm TSC\sim2.5~Gyr$) in \texttt{M9S0-D} and \texttt{M9S0}, respectively, while the sub-cluster is no longer trackable after its second pericenter passage in the SIDM runs ($\rm TSC<2~Gyr$).

In all simulations, the overall mass distribution is consistent regardless of self-interactions, presenting subhalos at comparable locations. The difference becomes visible close to the cluster center, where dark matter density is high. When compared with the corresponding CDM simulations, both the DMO and MHD simulations with the SIDM model present smooth dark matter density distributions with a thermalized core. This demonstrates that the self-interaction implemented in the simulations acts as designed, modifying the dense regions while preserving the successful prediction of the $\Lambda$CDM paradigm on large-scale structures \citep[e.g.,][]{2013MNRAS.430...81R}. 

\begin{figure}
 \centering
\includegraphics[width=0.9\columnwidth]{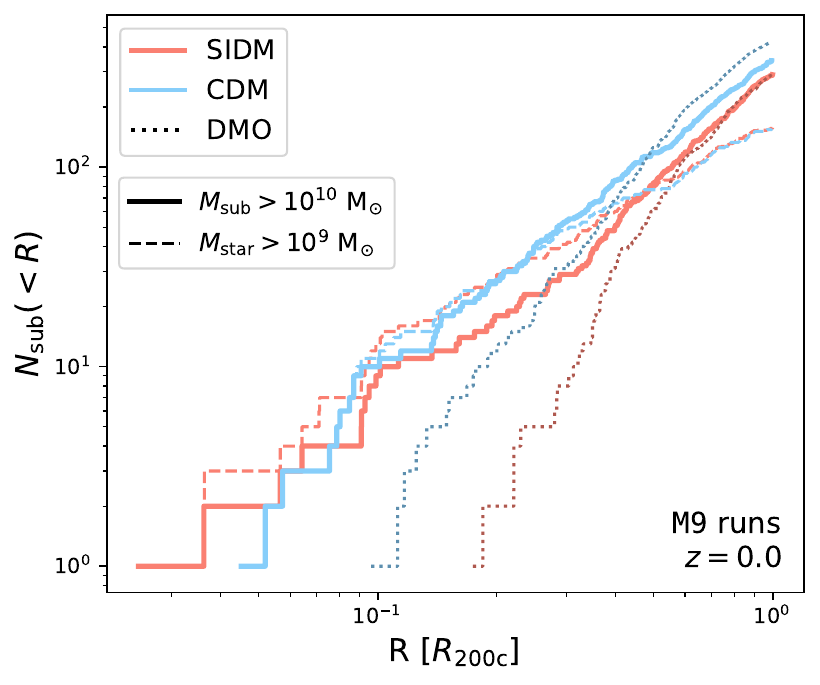}
 \caption{Cumulative number of subhalos within $R_{\rm 200c}$ of the massive cluster merger runs at $z=0.0$. The blue and red lines depict the CDM (\texttt{M9S0}) and the SIDM with $\sics=1\rm~cm^{2}~g^{-1}$ (\texttt{M9S1}) runs, respectively. The subhalos are selected based on their total mass ($M_{\rm sub}>10^{10}\rm~M_{\odot}$; solid) and stellar mass defined within two times the half-mass radius ($M_{\rm star}>10^{9}\rm~M_{\odot}$; dashed). The numbers of dark matter subhalos from the DMO simulations are overlaid for comparison (dotted). The number of dark matter subhalos is smaller in the SIDM runs, with the difference becoming larger toward the cluster center, while the number of stellar subhalos remains comparable between the two runs.
 }
 \label{fig:SubhaloNum}
\end{figure}

From the contour peaks in Figure~\ref{fig:DMmap_SIDM}, we can also see that the number of subhalo mass peaks decreases with self-interaction. Figure~\ref{fig:SubhaloNum} presents the cumulative number of subhalos, which shows a larger difference between the CDM and SIDM runs toward the cluster center. In the DMO simulations, compared to the \texttt{M9S0-D} run, the number of subhalos within $R_{\rm 200c}$ in \texttt{M9S1-D} with $M_{\rm sub}>10^{10}~M_{\odot}$ and $>10^{11}~M_{\odot}$ decreases by $32\pm5\%$ and $30\pm16\%$, respectively, assuming Poisson noise. These decrements increase to $51\pm8\%$ and $55\pm25\%$ within $0.5R_{\rm200c}~(\mytilde1\rm~Mpc)$, and no $10^{10}~M_{\odot}$ subhalo remains within $\mytilde0.2R_{\rm 200c}$ in \texttt{M9S1-D}, where the interaction becomes more active. The MHD simulations show a similar trend. The \texttt{M9S1} run contains $15\pm7\%$ and $30\pm15\%$ fewer subhalos within $R_{\rm 200c}$ than the \texttt{M9S0} run for $M_{\rm sub}>10^{10}~M_{\odot}$ and $>10^{11}~M_{\odot}$, respectively, and the decrements increase to $30\pm10\%$ and $47\pm23\%$ within $0.5R_{\rm200c}$. In contrast to the DMO runs, multiple $10^{10}~M_{\odot}$ subhalos remain within $0.1R_{\rm 200c}$ in both the CDM and SIDM runs. For $M_{\rm sub}>10^{10}~M_{\odot}$, the impact of self-interaction on the subhalo number appears weaker than in the DMO runs, possibly due to the additional potential provided by stellar components. For $M_{\rm sub}>10^{11}~M_{\odot}$, the decrements are consistent with the DMO runs within the uncertainties.

On the other hand, self-interactions have a different impact on the stellar mass content. As presented in the bottom row of Figure~\ref{fig:DMmap_SIDM}, the overall ICL morphology is comparable in the \texttt{M9S0} and \texttt{M9S1} runs.  The ICL extends preferentially along the collision axis, with its elongation matching that of the dark matter distribution. When subhalos are counted by their stellar mass defined within two times the half-mass radius, the numbers with $M_{\rm star}>10^{9}~M_{\odot}$ and $>10^{10}~M_{\odot}$ within $R_{\rm200c}$ and $0.5R_{\rm200c}$ are consistent between the \texttt{M9S0} and \texttt{M9S1} runs (see the dashed lines in Figure~\ref{fig:SubhaloNum}).\footnote{Unlike Figure~\ref{fig:SubhaloNum}, the stellar mass distribution in Figure~\ref{fig:DMmap_SIDM} shows more subhalo-like peaks in the SIDM run. As the map shows the ICL population that excludes stellar particles bound to subhalos, these peaks trace weakly bound stars of disrupted subhalos that are assigned to the ICL.} As the number of dark matter subhalos decreases with self-interactions, this implies that SIDM actively disrupts dark matter subhalos in the cluster environments, while their stellar components are mostly left unaffected \citep[e.g.,][]{2025A&A...697A.213D}. We note that this indicates that the number of galaxies predicted based on dark matter halos from the SIDM-only simulations may overestimate the impact of self-interaction on the galaxy mass function. However, the quantitative comparison of the stellar content should be interpreted with caution, as the subgrid models are calibrated based on the CDM models.

\begin{figure}
 \centering
\includegraphics[width=0.9\columnwidth]{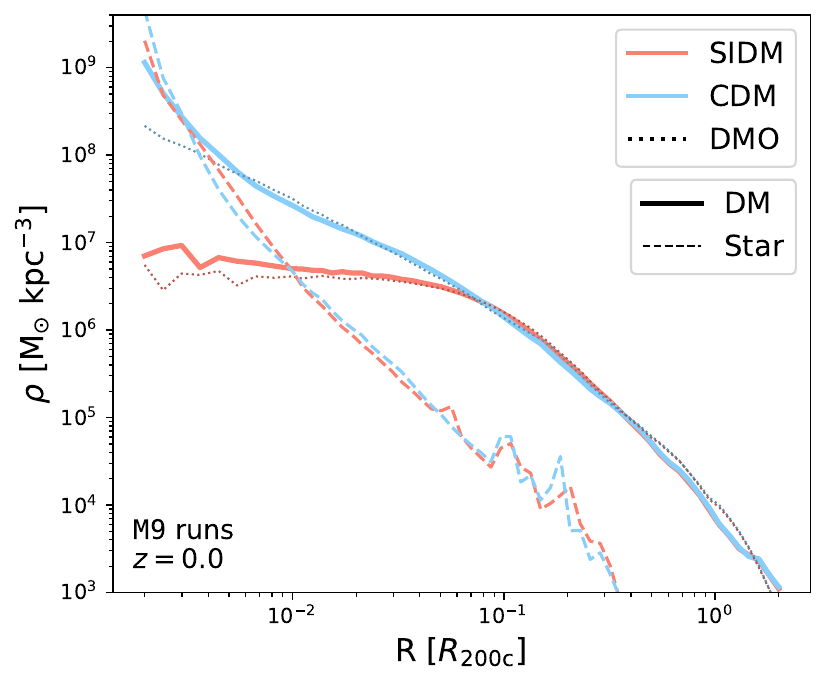}
 \caption{Dark matter (solid) and stellar density profiles (dashed) of the \texttt{M9S0} and \texttt{M9S1} runs at $z=0.0$. The collisionless and self-interacting dark matter models with $\sics=1\rm~cm^{2}~g^{-1}$ are marked with blue and red lines, respectively. Dark matter density profiles from the DMO simulations are overlaid for comparison. Dark matter self-interactions have generated a thermalized core in the dark matter distribution, while the stellar density profile presents comparable slopes across all radii. The MHD runs show a higher dark matter density than their DMO counterparts in the innermost region, which can be attributed to the additional gravitational potential of the baryons.
 }
 \label{fig:SIDMhalo}
\end{figure}

\subsection{Density profiles}
\label{sec:SIDM_profile}

Figure~\ref{fig:SIDMhalo} further compares the dark matter and stellar density profiles between the simulations. We can clearly identify the isothermal core of the dark matter halo in the SIDM runs. The SIDM runs present a dark matter density profile slope consistent with the CDM runs in the cluster outskirts. Within $0.08R_{\rm 200c}~(\mytilde150\rm~kpc)$, the dark matter density profiles in the SIDM runs become flat and so present an order of magnitude lower density than the CDM runs at the center. We note that the dark matter density is marginally larger in the SIDM runs at $\mytilde0.1R_{\rm 200c}$, and so the enclosed dark matter mass within $0.2R_{\rm 200c}$ is comparable between CDM and SIDM runs. The stellar density profile, on the other hand, is relatively similar throughout all radii. The \texttt{M9S1} run presents a lower stellar density in the central region within $<0.03R_{\rm 200c}$, which can be interpreted as the result of weakly bound stellar particles in the shallower gravitational potential and modified star formation histories \citep[e.g.,][]{2014MNRAS.444.3684V}. The total stellar mass content within $R_{\rm 200c}$ at $z=0$ is $1.1\times10^{13}~M_{\odot}$ in the \texttt{M9S0} and $9.9\times10^{12}~M_{\odot}$ in the \texttt{M9S1} run. 

Under the same dark matter models, the inclusion of baryons increases the dark matter density. In the CDM runs, \texttt{M9S0} presents higher dark matter density than the \texttt{M9S0-D} run at the innermost bin with $r< 0.005R_{\rm 200c}~(\sim10\rm~kpc)$.\footnote{We note that this radius is larger than both the softening length $\mytilde1.4\rm~kpc$ and the convergence radius \citep[$\mytilde8\rm~kpc$,][]{2003MNRAS.338...14P,2008MNRAS.391.1685S} of the \texttt{M9S0} halo. The number of dark matter particles in this bin is $\mytilde10^4$.} In the SIDM runs, \texttt{M9S1} presents a marginally higher core density than \texttt{M9S1-D}. These differences can be attributed to the additional gravitational potential provided by the stellar core, as the variation appears when the stellar density exceeds the dark matter density \citep[e.g.,][]{2026A&A...708A..47D}. Note that the ICM density in the \texttt{M9S0} and \texttt{M9S1} runs presents a flat profile within $\mytilde0.1R_{\rm 200c}$, and so does not exceed either the dark matter or stellar mass density in the central region.

\begin{figure}
 \centering
 \includegraphics[width=0.9\columnwidth]{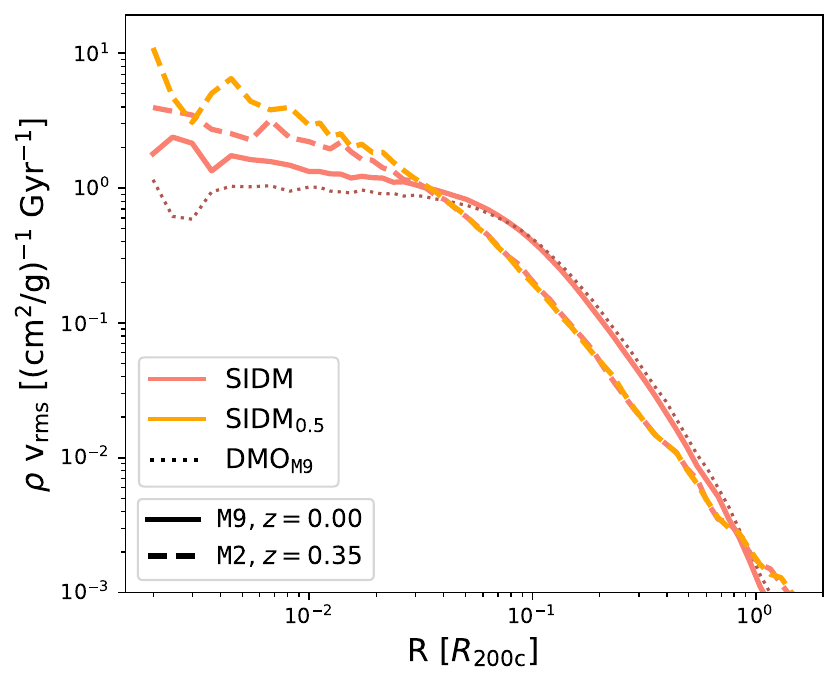}
 \caption{Expected interaction rate per cross-section in the SIDM runs of the massive (\texttt{M9} runs, solid) and low-mass cluster runs (\texttt{M2} runs, dashed). The red lines show the SIDM runs with $\sics=1\rm~cm^{2}~g^{-1}$, and the orange line shows the SIDM run with $\sics=0.5\rm~cm^{2}~g^{-1}$ in the low-mass cluster (\texttt{M2S05}), labeled SIDM$_{0.5}$. The dotted line shows the measurement from the DMO simulation of the massive cluster. The low-mass cluster is sampled at $z=0.35$ ($M_{\rm 200c}=4.9\times10^{13}\rm~M_{\odot}$), when the cluster is in a relatively relaxed phase. The flat profile that mimics the density profile suggests an isothermal core with a constant velocity dispersion. Both the massive and low-mass cluster runs show a flat profile where the expected interaction rate is significant within the cluster dynamical timescale.}
 \label{fig:Rate}
\end{figure}
Mass maps and density profiles clearly demonstrate the role of self-interactions in redistributing the dark matter particles. To examine how often particles need to interact to create these differences, we estimate the interaction rate as $v_{\rm rms}(\sics)\rho_{\rm DM}$, where $v_{\rm rms}$ is the velocity dispersion of the dark matter particles within each spherical shell. Figure~\ref{fig:Rate} presents the expected interaction rate per cross-section of the dark matter halo in the \texttt{M9} and \texttt{M2} runs. We select the snapshot where the merger is complete to minimize the impact of bulk motion of the cluster merger on the velocity dispersion measurement. As the major merger does not complete at $z=0$ in the low-mass cluster merger, we analyze the cluster at $z=0.35$, $\mytilde2.2~\rm Gyr$ before the first pericenter passage, when the cluster is in a relatively isolated state before the merger. 

In the \texttt{M9S1} run, the halo presents a flat interaction rate per cross-section in the central region within $\mytilde0.08R_{\rm 200c}$, which is consistent with the radius where the density profile flattens. This implies that the velocity dispersion of dark matter within this central region is consistent across the radius, such that the dark matter halo is isothermal. Based on this estimate, we can predict that the dark matter halo develops a flat density slope when the self-interaction rate reaches $\sim1-2~\rm Gyr^{-1}$ \citep[e.g.,][]{2013MNRAS.430...81R}. Similar to the density profile, the expected interaction rate is higher in \texttt{M9S1} than in the \texttt{M9S1-D} run. Within $0.02R_{\rm 200c}$, the median interaction rate per cross-section is $\mytilde40\%$ higher in the MHD run than its DMO counterpart. In the same region, we find that the velocity dispersion of the dark matter particles is only greater by $\mytilde4\%$ in the \texttt{M9S1} run, indicating the higher expected interaction rate in the MHD simulation is mainly driven by the enhanced density, which, as discussed in Figure~\ref{fig:SIDMhalo}, can be attributed to the additional potential from the baryonic component.

The interaction rate per cross-section presents a similar trend in the \texttt{M2} runs, showing a flat profile when the density profiles start to deviate from their CDM run counterparts. We find that the expected interaction rate is comparable between \texttt{M2S1} and \texttt{M2S05}. When multiplied by the assumed $\sics$, both \texttt{M2S1} and \texttt{M2S05} runs predict an interaction rate of $\sim3~\rm Gyr^{-1}$ at the cluster center. This interaction rate is similar to that of the \texttt{M2S1-D} and \texttt{M9} runs of $\mytilde2\rm~Gyr^{-1}$. As the expected interaction rate is comparable across the simulations, we interpret that the morphological variations (Figure~\ref{fig:DMmap_SIDM}) and density profile variations (Figure~\ref{fig:SIDMhalo}) under the SIDM paradigm appear when the dark matter interaction can occur within the system's dynamical timescale.

\begin{figure}
 \centering
 \includegraphics[width=0.9\columnwidth]{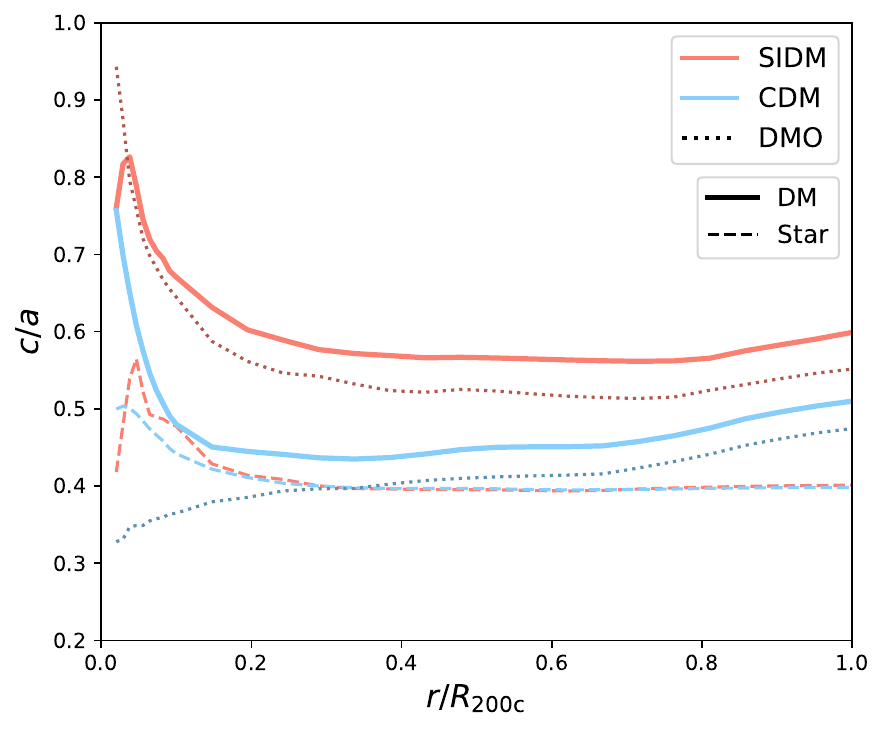}
 \caption{Ellipticity profile at $z=0$ defined using the ratio of the minor ($c$) and major ($a$) axes, for both dark matter (solid) and stellar (dashed) mass distributions of the massive cluster. The blue and red lines depict the CDM (\texttt{M9S0}) and the SIDM (\texttt{M9S1}) runs, respectively. The ellipticity of the dark matter from the DMO simulations (dotted) is overlaid for comparison. The axis ratios are calculated using particles enclosed within each radius that are identified by \texttt{SUBFIND} as bound to the main cluster. The dark matter distribution is rounder in the SIDM runs, while we do not find a large difference in the stellar mass distribution or in the dark matter distribution of the innermost region ($\mytilde0.03R_{\rm 200c}$).
 }
 \label{fig:Axis_ratio}
\end{figure}
\subsection{Halo shape}
\label{sec:SIDM_halo_shapes}

As discussed in previous studies \citep[e.g.,][]{2013MNRAS.430..105P, 2018MNRAS.474..746B, 2019MNRAS.488.3646R}, the dark matter halos in Figure~\ref{fig:DMmap_SIDM} present a more spherical morphology in the SIDM simulations. Following \citet{2019MNRAS.488.3646R}, we estimate the ellipticity by iteratively calculating the inertia tensor of particle distributions. At each radius $r$, we gather the particles within the radius and calculate the eigenvalues of the reduced inertia tensor ($\tilde{I}_{ij}$), 
\begin{equation}
    \tilde{I}_{ij} =\sum_n \frac{x_{\rm i,n} x_{\rm j,n} m_n}{r_{\rm n}^2 } / \sum_n  m_{\rm n},
\end{equation}
where the position of the nth particle is described as $(x_{\rm 1,n}, x_{\rm 2,n}, x_{\rm 3,n})$ and $r_{\rm n}$ is the ellipsoidal radius defined as $r_{\rm n} = \sqrt{x_{\rm 1,n}^2/a^2 +x_{\rm 2,n}^2/b^2 +x_{\rm 3,n}^2/c^2}$. In the first calculation, we use $a=b=c=1$, and compute the coordinates along the eigenvector basis. Then, we repeat the computation using particles satisfying $r_n^3 < (a^2/bc)\,r^3$. We repeat the process until the difference in both axis ratios, $b/a$ and $c/a$, between two sequential iterations is smaller than $1\%$. At the smallest radius ($<0.03R_{\rm 200c}$ or $<60\rm~kpc$), all runs use $\gtrsim10^4$ particles for estimating the shape. For this analysis, we use the particles that are bound to the main cluster to avoid the contribution from infalling subhalos \citep[e.g.,][]{2022MNRAS.516.4543D}. 

\begin{figure*}
 \centering
 \includegraphics[width=2\columnwidth]{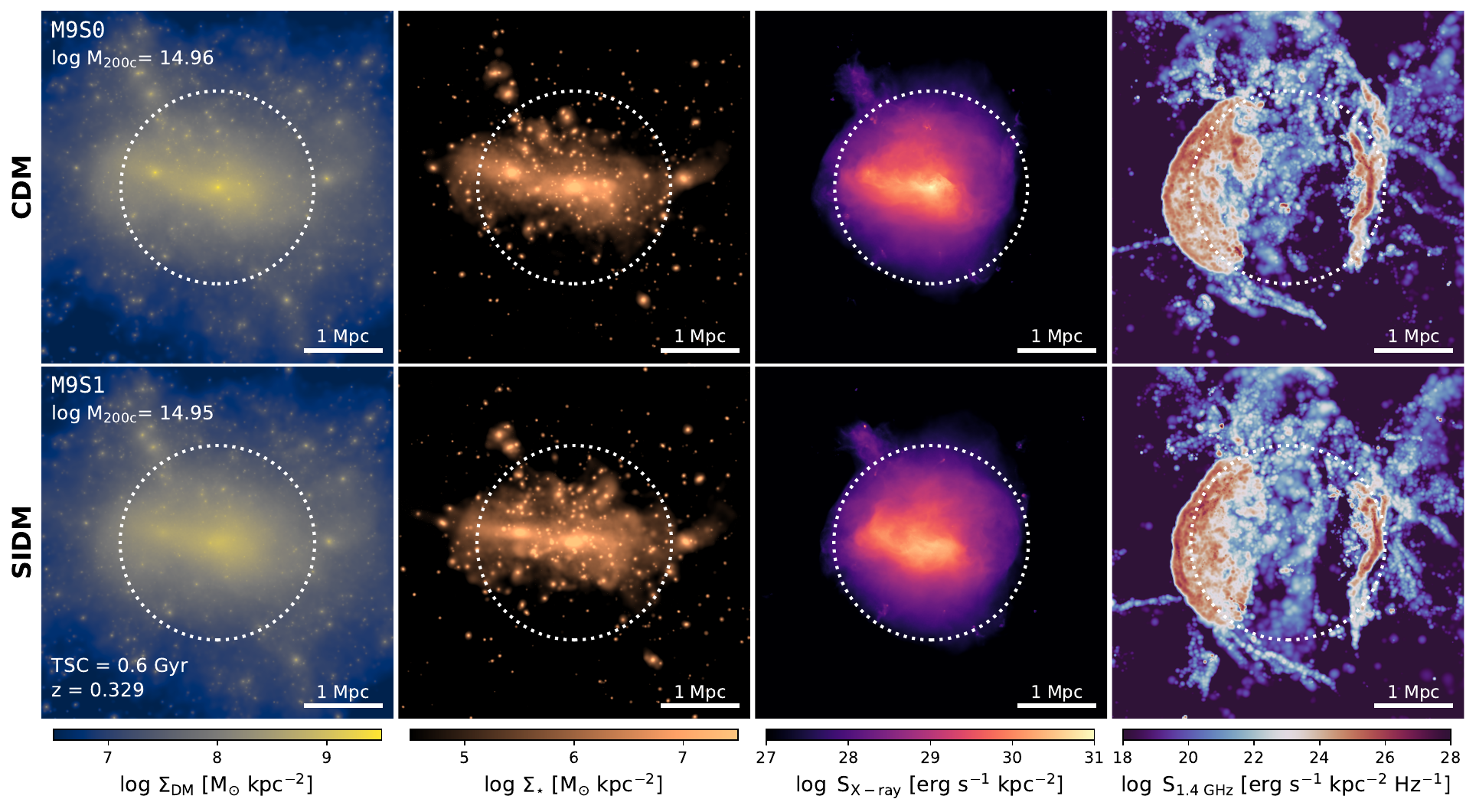}
 \caption{Multi-wavelength projection maps of the simulated cluster merger with collisionless (top, \texttt{M9S0}) and SIDM with $\sigma_{\rm T}/\rm m_{\chi}=1\rm~cm^2~g^{-1}$ (bottom, \texttt{M9S1}). From left to right, the columns display the projected dark matter column density, stellar column density, X-ray surface brightness, and radio surface brightness. The annotated circle marks $R_{500c}$ centered on the main cluster center. The snapshot corresponds to $\mytilde0.6\rm~Gyr$ after the first pericenter passage. The large-scale merger morphology is broadly consistent across dark matter models, while the detailed morphology shows variations.}
 \label{fig:mw_relic_zoo}
\end{figure*}

Figure~\ref{fig:Axis_ratio} shows the ellipticity, quantified by the axis ratio $c/a$, measured using the dark matter and stellar particles in the massive cluster at $z=0$. The dark matter halos in the SIDM runs are more spherical, with a larger $c/a$. The \texttt{M9S1} and \texttt{M9S1-D} runs have $c/a\sim[0.5,0.6]$, higher than their CDM counterparts \texttt{M9S0} and \texttt{M9S0-D} with $c/a\sim[0.4,0.5]$. This difference extends to the cluster outskirts, whereas it decreases toward larger radii in previous cosmological simulations \citep[e.g.,][]{2019MNRAS.488.3646R,2026A&A...706A.340G}. We infer that this is related to the major merger, which elongates the mass distribution along the collision axis. In contrast, the axis ratio of the stellar mass distribution is comparable between the \texttt{M9S1} and \texttt{M9S0} runs, consistent with previous simulations \citep[e.g.,][]{2019MNRAS.488.3646R,2026MNRAS.551g1423M}. From the cluster outskirts to the inner region of $\mytilde0.2R_{\rm 200c}$, the stellar mass distribution has $c/a\sim0.4$ in both the CDM and SIDM runs, which is more elliptical than the host dark matter halo.

In the innermost region, on the other hand, we find that the shapes become spherical. Except for the dark matter in the \texttt{M9S0-D} run, dark matter and stellar particles present a more spherical distribution in the central region $<0.1R_{\rm 200c}$. Specifically, the difference between the dark matter halo axis ratio in the \texttt{M9S1} and \texttt{M9S0} runs decreases and shows a comparable ratio of $c/a\sim0.8$ at the centers. These results show that, although the dark matter self-interaction of $\sics=1\rm~cm^2~g^{-1}$ can modify the dark halo shapes, the difference in halo shape may vary with the radius \citep[but see also][]{2023MNRAS.521.3172R}.

\subsection{Multi-wavelength observables of cluster major mergers}
\label{sec: SIDM_Merger}

Our re-simulated clusters experience a major merger between two clusters. The cluster in the \texttt{M9} run experiences a major merger with a mass ratio of $\mytilde1:3$, in a Bullet-like merger configuration, and a minor merger event at a higher redshift. The cluster in the \texttt{M2} run experiences a major merger with a mass ratio of $\mytilde1:2$, with a smaller collision velocity and a larger pericenter separation than the \texttt{M9} run. 

Figure~\ref{fig:mw_relic_zoo} presents the multi-wavelength observables of the \texttt{M9S0} and \texttt{M9S1} runs at $\mytilde0.6\rm~Gyr$ after the first pericenter passage of the major merger. The overall morphology is similar between the two runs. The major merger occurs along the x-axis of the image, presenting the dark matter and stellar mass distributions as extending along the merger axis. The X-ray surface brightness map shows an unrelaxed morphology, with an X-ray tail extending from each cluster toward its colliding counterpart. In radio, two giant arc-shaped emission features appear in front of each cluster, which can be classified as double radio relics. These double radio relics have a luminosity ratio of $\mytilde3$, comparable to the mass ratio of the collider. 
 
The detailed morphologies vary between the runs during the merger as well. Similar to the cluster at $z=0$, the dark matter distribution near the cluster center is more diffuse in the SIDM run, with fewer subhalo-associated density peaks present in the innermost regions. The stellar mass distribution remains similar between the runs, presenting dense peaks and a large number of stellar subhalos. On the other hand, we find that the ellipticity of the stellar mass distribution in the SIDM run is more elongated than that of the CDM run. We repeat the axis ratio calculation from \textsection\ref{sec:SIDM_halo_shapes}, now combining the ICL populations of the two colliding clusters. The axis ratio $c/a$ of the stellar mass distribution, measured at the main-cluster and sub-cluster centers within $0.1R_{\rm 200c}~(\mytilde180\rm~kpc)$, is 0.73 and 0.71 in \texttt{M9S0}, while the ratios are 0.65 and 0.45 in \texttt{M9S1}. On a larger scale, a narrow stellar stream stretching beyond $R_{\rm 500c}$ from the sub-cluster can be identified in the SIDM cluster merger. Since the ICL axis ratios of the two runs converge to similarly small values at $z=0$, the difference in the ICL shape appears to change with time, and the merger likely elongates the ICL in both runs.

The features associated with the intracluster plasma also show variations in their detailed morphology. Overall, the X-ray surface brightness presents a similar extension to the dark matter, while the presence of the dense core in the main cluster and the morphology of the X-ray surface brightness discontinuity vary with self-interactions. As the sub-cluster does not present a dense X-ray core before the collision, the high-velocity, head-on collision has disrupted the X-ray distribution in both \texttt{M9S0} and \texttt{M9S1} runs, making dark matter halo-X-ray dissociation unclear. Nevertheless, at an earlier merger phase, both the BCG and dark halo centroid of the sub-cluster present a spatial offset from the bright X-ray regions. When we measure the separation of the cluster center to the X-ray surface brightness contour of $10^{30.5}\rm~ erg~s^{-1}~kpc^{-2}$ level at $\rm TSC\sim0.2\rm~Gyr$\footnote{This is the X-ray surface brightness level of the sub-cluster core at the pre-merger phase.}, the minimum separation from the X-ray contour to the sub-cluster BCG is $\mytilde100\rm~kpc$ in both \texttt{M9S0} and \texttt{M9S1} runs, comparable to the dissociation reported in the Bullet Cluster. Thus, although dark matter with a large $\sics$ may behave like ICM in dissociation \citep[e.g.,][]{2004ApJ...606..819M}, self-interaction driven by $\sics=1\rm~cm^{2}~g^{-1}$ can create dark halo-X-ray dissociation at a level comparable to that in the CDM mergers.

\begin{figure}
 \centering
 \includegraphics[width=\columnwidth]{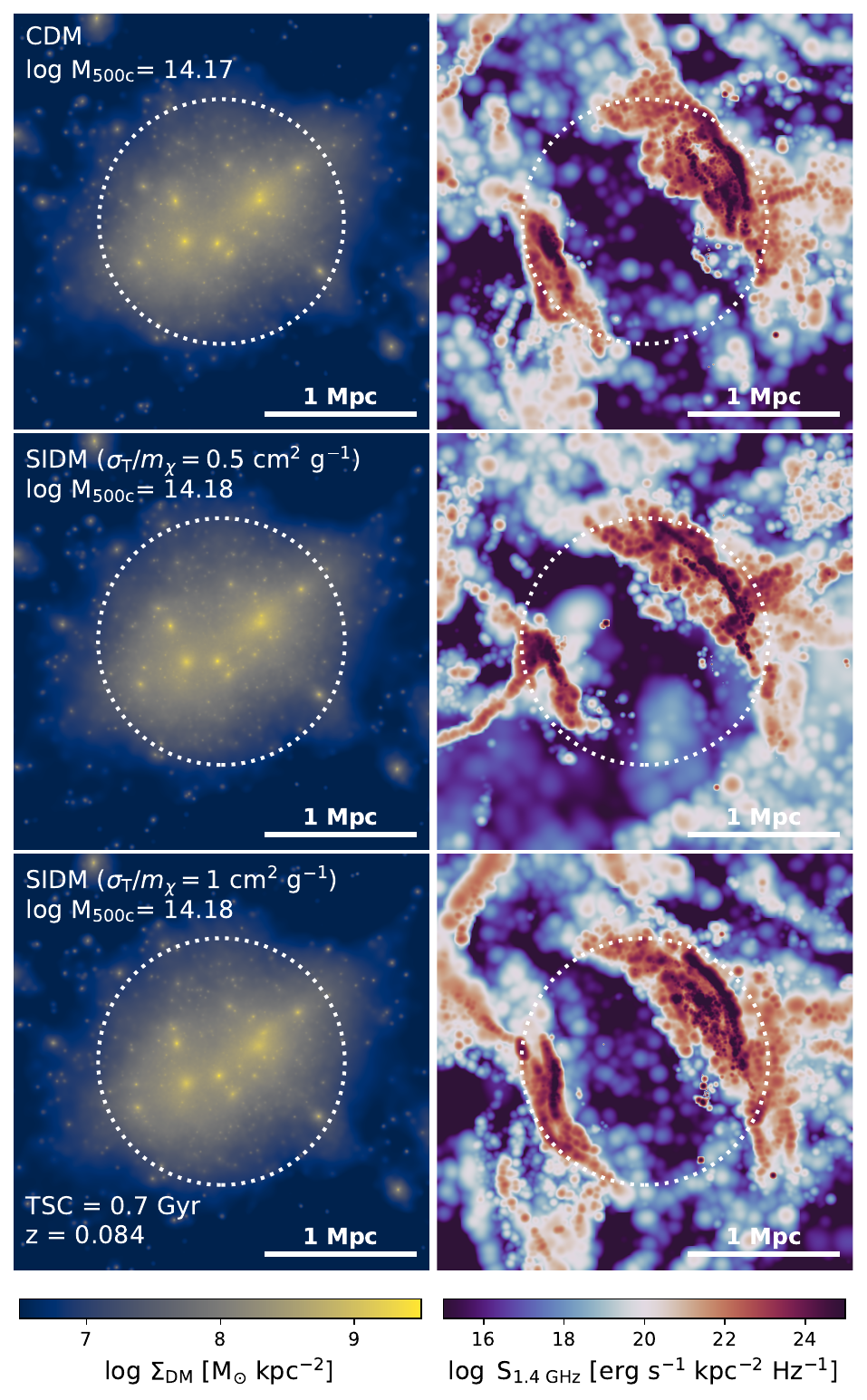}
 \caption{Same as Figure~\ref{fig:mw_relic_zoo}, but for the low-mass cluster merger in the \texttt{M2S0} (top), \texttt{M2S05} (middle), and \texttt{M2S1} (bottom row) runs. Similar to the massive cluster merger, large-scale morphology is consistent across the runs. However, the eastern radio relic in the \texttt{M2S05} run fades more quickly than in the other two runs, possibly due to stochastic effects of self-interaction in galaxy formation.}
 \label{fig:mw_relic_zoo_low}
\end{figure}

The radio relic morphology varies with dark matter self-interactions. In Figure~\ref{fig:mw_relic_zoo}, we can find that the eastern relic has a brighter edge in the \texttt{M9S1} run, while the western relic has a larger extent in the \texttt{M9S0} run. To further explore the connection between diffuse radio emission and self-interaction, we compare the radio surface brightness maps from the three low-mass cluster merger runs, \texttt{M2S0}, \texttt{M2S05}, and \texttt{M2S1}, in Figure~\ref{fig:mw_relic_zoo_low}. As before, the collision axis is aligned with the x-axis of the image. Similar to the massive cluster merger, the overall dark matter and radio surface brightness morphologies are similar across the three runs. Nevertheless, radio emission exhibits variations that do not scale with the cross-section. Specifically, in the \texttt{M2S05} run, the secondary relic propagating toward the east of the image is less spatially extended and shows substructure along the x-axis, compared to the \texttt{M2S0} and \texttt{M2S1} runs. At this snapshot, the integrated radio luminosities are comparable among all three simulations. However, at a late merger phase, the radio relics in the \texttt{M2S05} run fade more rapidly, and their luminosity drops below our threshold at  $\mytilde1~\rm Gyr$ after pericenter passage, $\mytilde0.5~\rm Gyr$ earlier than in the other two runs. 

\begin{figure}
\centering
\includegraphics[width=0.94\columnwidth]{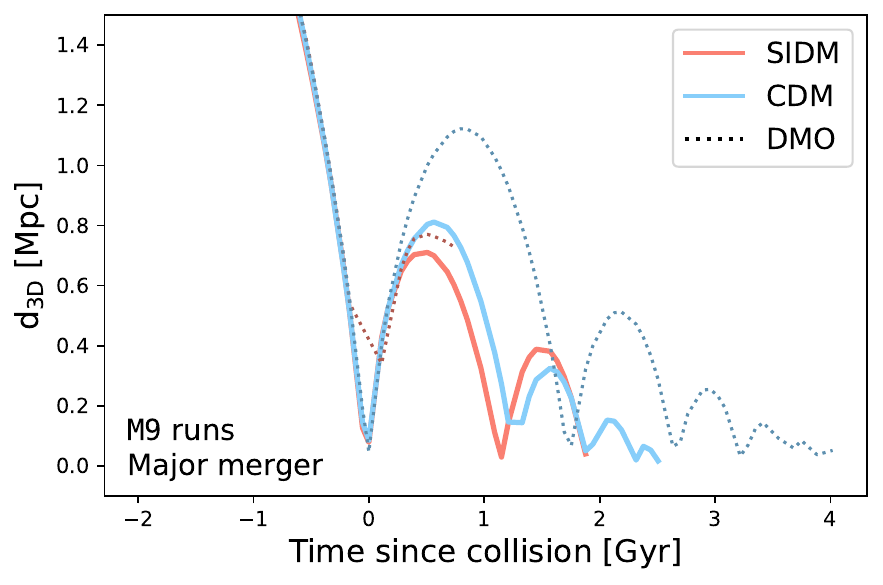}
\includegraphics[width=0.94\columnwidth]{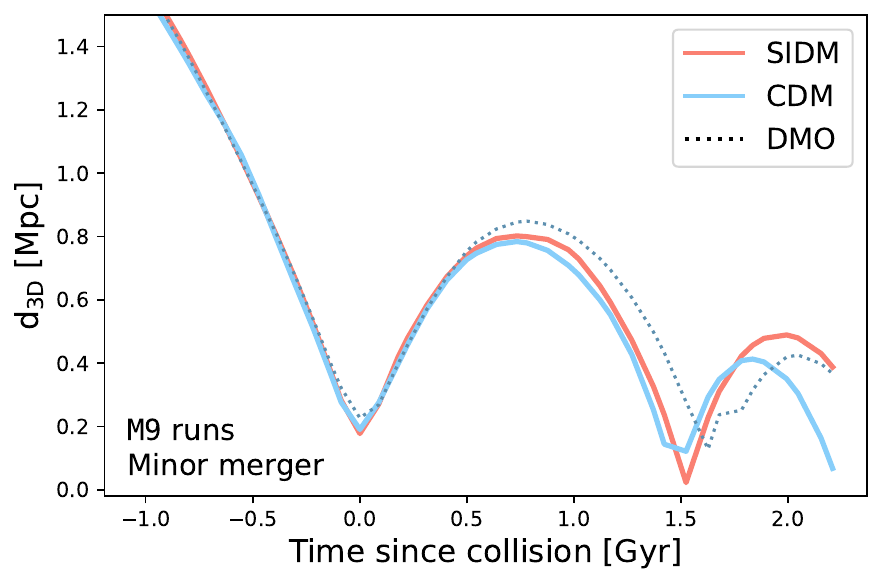}
\includegraphics[width=0.94\columnwidth]{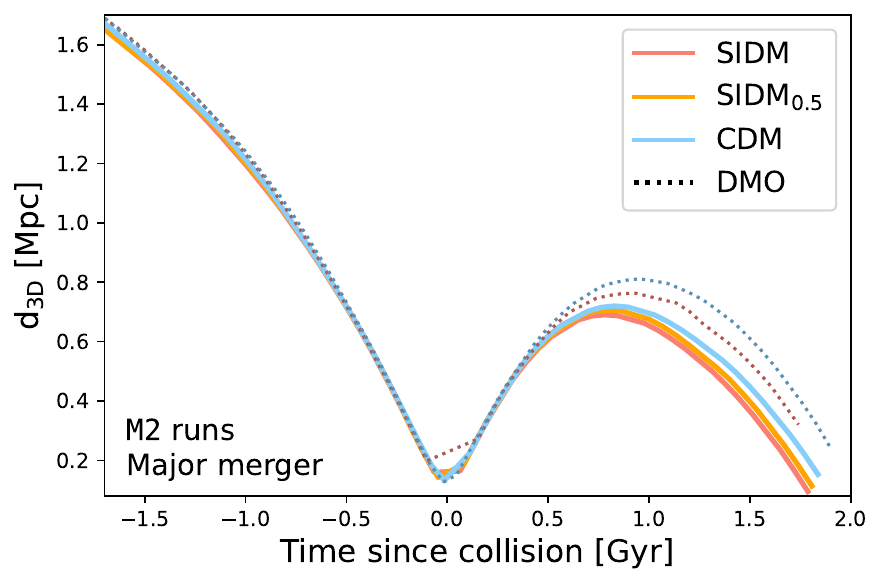}
\caption{Time evolution of the halo separation in the \texttt{M9} major merger (top), the \texttt{M9} minor merger (middle), and the \texttt{M2} major merger (bottom row). Line color indicates the dark matter model: red, orange, and blue lines represent the SIDM model with $\sics=1\rm~cm^2~g^{-1}$, $\sics=0.5\rm~cm^2~g^{-1}$, and the CDM model, respectively. Solid lines show the measurements from the MHD simulations, and dashed lines show those from the DMO simulations. In both massive and low-mass cluster mergers, self-interaction modifies the halo orbit after the first pericenter passage ($\rm TSC=0$), while the effect of self-interaction may vary with the mass ratio and the merger phase. }
\label{fig:Halo_orbits}
\end{figure}

We attribute these differences in relic properties to environmental variations across the runs. At $\mytilde1~\rm Gyr$ after pericenter passage, when the secondary relic in the \texttt{M2S05} run approaches the detection threshold, the median Mach number of its shock cells is $\mytilde3.3\pm0.7$, lower than in the \texttt{M2S0} ($\mytilde4.3\pm1.2$) and the \texttt{M2S1} runs ($\mytilde3.8\pm1.1$). Although these differences lie within the standard deviations, radio emissivity is a sensitive function of Mach number in the weak-shock regime \citep[$\mathcal{M} \lesssim4$;][]{2015ApJ...809..186K}, such that a low Mach number distribution with lower variation is expected to yield fainter emission. In addition, by tracing the time evolution, we identify an AGN feedback event from a neighboring subhalo in the \texttt{M2S05} run. This energy injection modifies the upstream environment in \texttt{M2S05} and boosts the median upstream temperature of the secondary relic by $\mytilde50\%$ relative to the other two runs. In the \texttt{M9} runs, although the double radio relic morphology varies between the CDM and SIDM runs, the relics do not fade more quickly in either run, and the shock properties are consistent between the \texttt{M9S0} ($\mathcal{M}\sim3.5\pm1.1$) and \texttt{M9S1} ($\mathcal{M}\sim3.4\pm1.2$) runs. These differences do not scale with self-interaction and may reflect the stochastic divergence of galaxy formation between otherwise identical runs, rather than a systematic effect of self-interaction \citep[e.g.,][]{2014MNRAS.444.3684V,2019ApJ...871...21G}. With a single realization per model, we cannot separate stochastic variation from a systematic effect of self-interaction. Simulations with varying merger configurations, SIDM models, and galaxy formation models are needed to understand this behavior.

\subsection{Orbit of merging clusters}
\label{sec:SIDM_orbit}

\begin{figure}
\centering
\includegraphics[width=1\columnwidth]{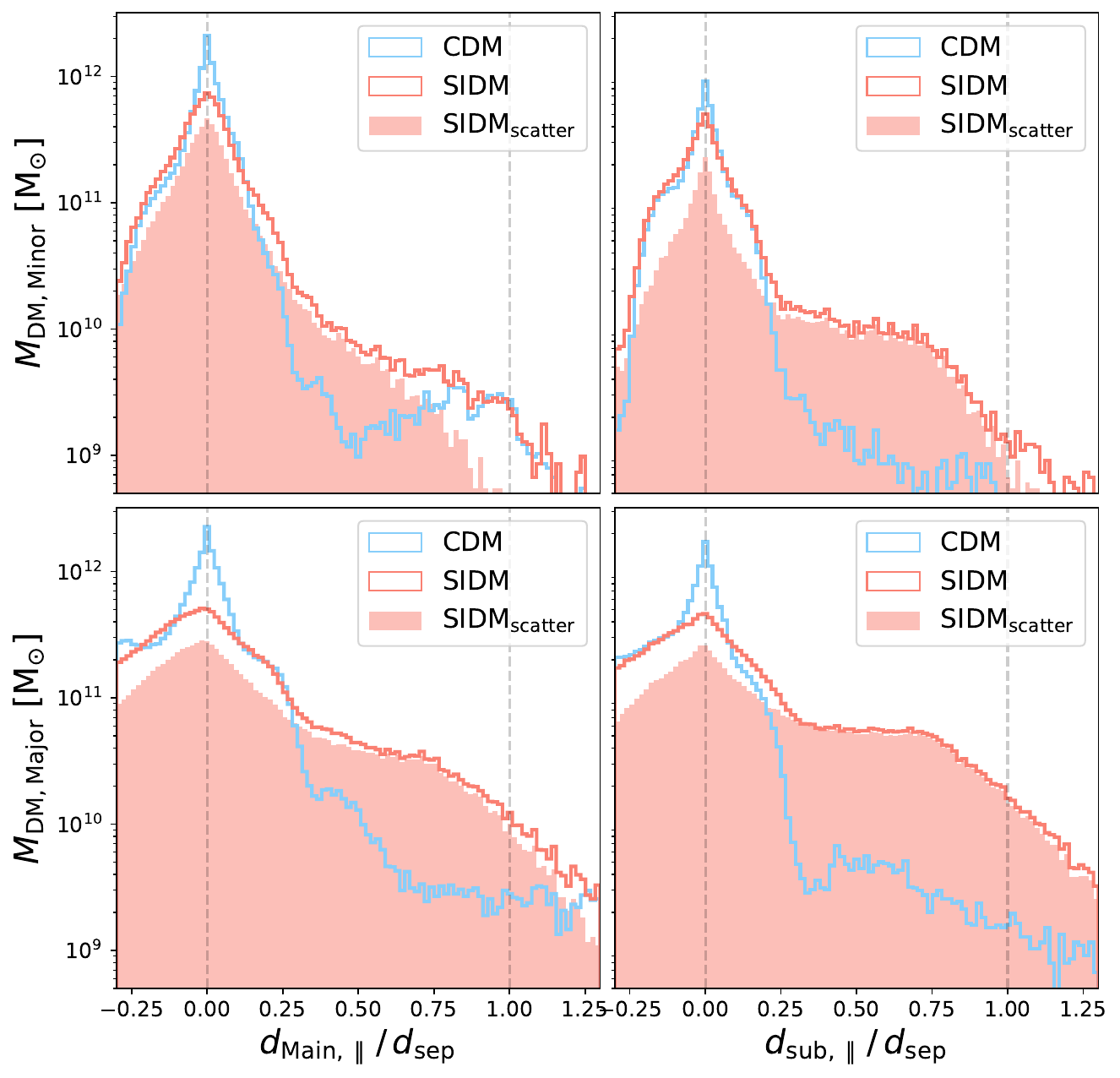}
\caption{Spatial distribution of dark matter particles after the collision, in the main (left column) and sub-cluster (right column), during the minor (top row) and major merger (bottom row) of the \texttt{M9} runs. The positions of the particles $0.4~\rm Gyr$ after the pericenter passage are presented, where the particles are selected within $100\rm~ckpc/h$ of each cluster center $0.4~\rm Gyr$ before the pericenter passage. The spatial distribution of particles is defined along the axis connecting the two clusters and normalized by the cluster separation, so that the target cluster and its companion cluster positions are fixed at 0 and 1, respectively. The blue and red histograms show the \texttt{M9S0} and \texttt{M9S1} mass distributions, respectively, and the filled red histogram shows the \texttt{M9S1} particles that experienced at least one self-interaction between the two timesteps separated by $\mytilde0.8\rm~Gyr$. A significant excess of dark matter mass is distributed between the clusters after the major merger by the particles that experience scattering, while the excess is insignificant during a minor merger.
}
\label{fig:DMtail}
\end{figure}

Figure~\ref{fig:Halo_orbits} presents the time evolution of the halo separation for the cluster merger events in the massive and low-mass clusters. We measure the halo separation using the halo center separations from the \texttt{SUBFIND} catalogue, which defines the halo position as the location of the particle with the minimum potential. Unlike the MHD simulations, where the gravitational potential of the baryonic component prevents the sub-cluster from being fully disrupted, the SIDM-only simulations lack this anchor. As a result, the dark matter halo of the sub-cluster cannot be identified at the late merger phase or even after the first pericenter passage in the SIDM-only runs. Thus, we exclude the snapshots where the sub-cluster is disrupted and can no longer be identified. We also stop recording the halo separation in the minor merger of \texttt{M9} runs after its second apocenter passage, as the first pericenter passage of the major merger occurs at $\rm TSC \sim 2.5\rm~Gyr$ from the first pericenter passage of the minor merger. 

The evolution of halo separation shows no variation before the pericenter passage, indicating a comparable collision velocity between the CDM and SIDM runs. After the collision, the halo separation is reduced at the same TSC, and the clusters reach their first apocenter earlier with the implementation of the SIDM model in both the massive and low-mass cluster mergers, and in the MHD and DMO simulations. The amount of decrement in the halo separation also depends on $\sics$, as we find that the halo separation of the \texttt{M2S05} run lies between those of the \texttt{M2S0} and \texttt{M2S1} runs. The halo separation evolves differently when comparing the MHD and DMO simulations. With the presence of ICM that dissociates from the dark halo during the pericenter passage, the ICM exerts additional gravitational drag on the clusters during their outgoing phases, making the cluster mergers in MHD runs reach apocenter at an earlier time at smaller halo separation. 

Interestingly, we find that the self-interaction has a different impact on the halo orbit during the minor merger and after the second pericenter passage. Unlike halo orbits in the major mergers, the first apocenter separation is larger with the presence of self-interaction during the minor merger (middle panel of Figure~\ref{fig:Halo_orbits}). A similar trend is presented after the second pericenter passage. In both the major and minor mergers of the \texttt{M9} cluster, the second apocenter separation in \texttt{M9S1} is larger than that in \texttt{M9S0}, and even larger than that of \texttt{M9S0-D} during the minor merger. Moreover, during the major merger, as the second collision begins from a free-fall at the first apocenter separation, a larger second apocenter separation in the \texttt{M9S1} run reflects that the effect of self-interaction on halo separation is opposite at this late merger phase. We note that a similar evolutionary trend is also presented in the idealized simulations \citep[e.g.,][]{2017MNRAS.469.1414K}.

We interpret the different impact as arising from how self-interactions exert a drag-like effect on the halo orbit in the rare-interaction regime \citep[e.g.,][]{2014MNRAS.437.2865K}. Due to rare interactions, only the particles that pass through the dense central regions experience interaction, while others effectively behave as collisionless particles. Scattered particles preferentially trail behind the motion of the clusters due to the random reorientation, which forms an overdensity along the collision axis and exerts an excess drag force on clusters during their outgoing phase. On the other hand, the dynamical friction scales with the mass of the infalling cluster and the local density of the background \citep[][]{1943ApJ....97..255C}, and so it can be weakened by the lower central dark matter density and by the active mass loss of the sub-cluster. As a result, we can expect that self-interactions can also hamper the halo deceleration after the pericenter passage. 

Figure~\ref{fig:DMtail} examines this by tracking the location of dark matter particles sampled from the cluster center. In minor and major merger of \texttt{M9} runs, we sample the dark matter particles within $100\rm~ckpc/h$ of each cluster $0.4~\rm Gyr$ before the pericenter passage and track their locations $0.4~\rm Gyr$ after the first pericenter passage. In the \texttt{M9S0} runs, the sampled particles reside in the cluster center at the post-merger phase. On the other hand, the dark matter particles in the \texttt{M9S1} run present an extended mass distribution towards its companion cluster. During the major merger, the mass distribution is extended with a relatively flat mass profile up to $\mytilde0.7 d_{\rm sep}$, which will exert a drag force on the clusters at their outgoing phase. We find a similar mass trend in the particles sampled from the sub-cluster during the minor merger. However, the mass distribution of the main cluster does not present a significant mass excess between clusters, and so the drag force in the outgoing phase will be insignificant compared to the major merger.

We further define the subsample of particles that experienced at least one self-interaction between these two snapshots. In both clusters, $\mytilde60\%$ and $\mytilde50\%$ of the selected particles experienced self-interaction during the major merger and minor merger, respectively. For comparison, we find that $\mytilde25\%$ of central particles experience self-interaction within $\mytilde0.8\rm~Gyr$ when repeating the process at a relatively relaxed state around $z=0$, suggesting that the interaction rate is doubled during the merger. 

\begin{figure}
 \centering
 \includegraphics[width=\columnwidth]{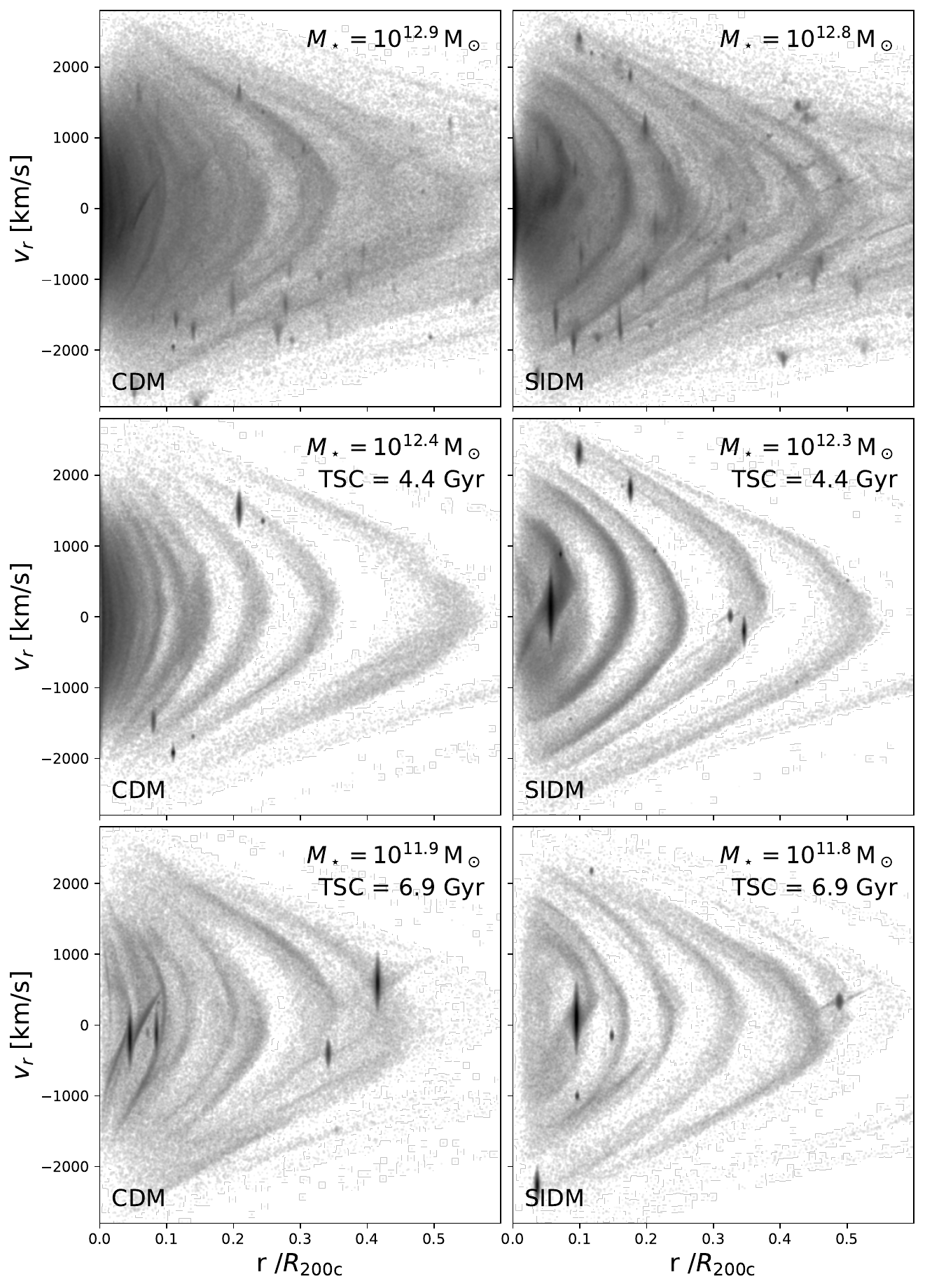}
 \caption{Radial velocity versus radius for the BCG+ICL stars in the \texttt{M9S0} (left) and \texttt{M9S1} runs (right). The top row shows all BCG+ICL stars identified with \textsc{subfind}. The middle and bottom rows show the subsample of these stars within 200\,kpc of the sub-cluster BCG in the major and minor mergers, respectively. The annotations mark the stellar mass of the intracluster light (top) and of the subhalo before infall (middle and bottom). The radius is normalized by $R_{\rm 200c}$, and the radial velocity is measured relative to the mass-weighted mean stellar velocity within $R_{\rm 200c}$. The color shows the stellar-mass-weighted surface density. The overdensities in the phase-space diagram, which could be observed as shell-like discontinuities in the ICL, can be associated with the stellar populations disrupted from the infalling BCGs and are expected to be more prominent under the SIDM scenario.}
 \label{fig:ICL_phase}
\end{figure}

When we plot the distribution of the scattered dark matter particles, the majority of the dark matter located between the two clusters have experienced self-interactions. Although a similarly large fraction of the dark matter sampled in the main cluster experiences scattering during the minor merger, its distribution remains mostly bound to the main cluster in the post-merger phase. This illustrates how self-interaction acts differently in major and minor mergers. During the major merger, the main and the sub-cluster move at a comparable velocity. The randomly reoriented motion of the scattered dark matter deviates from the bulk motion of the clusters, leaving a mass trail behind their motion. In the minor merger, however, the bulk velocity of the main cluster in the order of $\mytilde100\rm~km~s^{-1}$ is much smaller than the virial velocity of the system. As a result, the random reorientation acts as dynamical heating rather than producing the strong directional effect seen in the major merger, and so the main cluster does not develop the mass bridge that would otherwise exert a drag force and reduce the halo separation at the post-merger phase. A similar effect is expected during the second pericenter passage, as both collision velocity and the sub-cluster mass are reduced compared to the first pericenter passage, making the formation of mass bridges and, therefore, the excess drag force insignificant.

These results highlight the role of mass ratio and merger phase when using halo separation for testing SIDM scenarios. Minor mergers can produce prominent merger signatures, as in the Bullet Cluster \citep[e.g.,][]{2026ApJ..1005...28C} and PLCK G287.0+32.9 \citep[][]{2017ApJ...851...46F}, and the cluster mergers observed at their late merger phase could have experienced their second pericenter passage \citep[e.g.,][]{2025ApJ...984...24S}. Even with a large self-interaction cross-section of $\sics=1\rm~cm^{2}~g^{-1}$, halo separations measured from these samples may not present the signature of self-interaction, leading to underestimation of $\sics$. Thus, a careful sample selection that excludes minor mergers and those after the second pericenter passage is required when constraining $\sics$ based on the halo separation after the merger. 

\subsection{Signatures in the intracluster light}
\label{sec: SIDM_ICL}

Unlike dark matter particles, which show variations in density profile, subhalo mass function, and morphology, the stellar population does not present a large difference in these features. Such a different imprint of self-interaction on dark matter and the stellar population may provide an observable testbed of SIDM \citep[e.g.,][]{2025ApJ...987L..15C,2026ApJ..1008..193Y,2026MNRAS.551g1423M}. A detailed comparison of multi-wavelength features with the ICL population will be discussed in the future work, while here we briefly share the potential observational imprint of self-interaction on the ICL distribution.

Figure~\ref{fig:ICL_phase} presents the radius-radial velocity diagram of the BCG+ICL population in the \texttt{M9S0} and \texttt{M9S1} runs. This phase-space diagram is composed of four components, namely a smooth background population, stellar particles bound to the BCG at the center, an overdensity of ICL along the phase-space fold, and unmasked, extended stellar content from individual subhalos that appears as vertically extended substructures. Overall, the \texttt{M9S0} run is relatively smooth while the \texttt{M9S1} run still retains multiple substructures. In particular, the \texttt{M9S1} run hosts a larger population of dense, disk-like distributions, which can be interpreted as a weakly bound stellar population of subhalos under the SIDM paradigm \citep[e.g.,][]{2022MNRAS.511.5927S}. Both CDM and SIDM simulations show multiple phase-space folds across the cluster, while a larger stellar mass is concentrated along the same curve in the \texttt{M9S1} run. This difference is more significant in the central region, where the \texttt{M9S0} run shows a more continuous distribution of stellar streams, whereas the \texttt{M9S1} run has a sharper contrast (see the feature at $\mytilde0.17R_{\rm 200c}$ in Figure~\ref{fig:ICL_phase}). Since stars stripped onto a similar orbit populate the same phase-space fold, these overdensities will appear as shell-like substructures in the stellar mass distribution (see the ICL overdensity in the eastern outskirt of the \texttt{M9S1} run in Figure~\ref{fig:DMmap_SIDM}).

The shell-like structures can be considered as coherent orbits of the stellar halo disrupted from infalling galaxies \citep[e.g.,][]{2025ApJ...990...96J,2026ApJ...998...30J}. To better interpret this, the middle and bottom rows of Figure~\ref{fig:ICL_phase} track the stellar particles of the sub-cluster BCG in the major and minor mergers, selected within $200\rm~kpc$ of the sub-cluster centers. In both CDM and SIDM runs, the sampled stars form a phase-space fold at a similar location. However, close to the central region, the \texttt{M9S0} run presents a more continuous distribution of stellar streams, whereas the \texttt{M9S1} run presents a sharper contrast with respect to its surroundings. This difference is more significant for the stars sampled from the minor merger, which fell into the cluster $\mytilde7\rm~Gyr$ ago. The sharper contrast of the ICL shells may be interpreted as a result of weaker dynamical friction. The harmonic orbital motion inside the isothermal core of an SIDM cluster makes the dynamical friction less efficient \citep[][]{2006MNRAS.373.1451R}, and so the orbits of the stars stripped from the sub-cluster BCGs do not dissipate quickly in time. We note that a similar feature is presented in previous studies of SIDM galaxy clusters \citep[e.g., see Figure 4 in][]{2025A&A...697A.213D} or predicted in the MW-size halos under the SIDM paradigm \citep[e.g.,][]{2019MNRAS.485.1008B,2025MNRAS.542.1827F}.

It is noteworthy that the variation in the ICL shells under the SIDM paradigm extends beyond the radius at which the density profile deviates. Recent and upcoming optical band surveys sensitive to low-surface-brightness structures, including Euclid and the Rubin Observatory \citep[e.g.,][]{2022MNRAS.513.1459M}, could identify a hint of the SIDM with these ICL substructures \citep[see also][]{2020ApJS..247...43K}. However, the quantitative connection between dark matter self-interaction and the ICL shells remains unclear, as these features are also present in the CDM runs depending on their recent merger history \citep[e.g.,][]{2018MNRAS.480.1715P}. A statistical analysis of ICL shells using a larger sample of simulated SIDM clusters would be required to assess their role in testing the self-interacting nature of dark matter.

\section{Observable tracers of self-interaction in merging clusters}
\label{sec: SIDM_tracers}
Self-interaction leaves observable imprints on galaxy clusters, including the isothermal core of the dark matter halo, halo ellipticity, and surface-brightness discontinuities in the ICL, which could be used to constrain $\sics$. Similarly, self-interaction leaves imprints on the multi-wavelength observables during cluster mergers, while some features present a stochastic variation with $\sics$. In this section, we test the observational probes of SIDM proposed for merging clusters and examine the validity of the probes in constraining $\sics$. Specifically, we focus on the evolution of the double radio relic separation \citep[e.g.,][\textsection\ref{sec:shock_halo_tracer}]{2026arXiv260500093J}, the BCG-dark halo offset \citep[e.g.,][\textsection\ref{sec:offset}]{2004ApJ...606..819M,2015Sci...347.1462H,2024OJAp....7E..65R}, and the mass-to-light ratio \citep[M/L; e.g.,][\textsection\ref{sec:mlratio}]{2008ApJ...679.1173R}.

\begin{figure}
 \centering
 \includegraphics[width=\columnwidth]{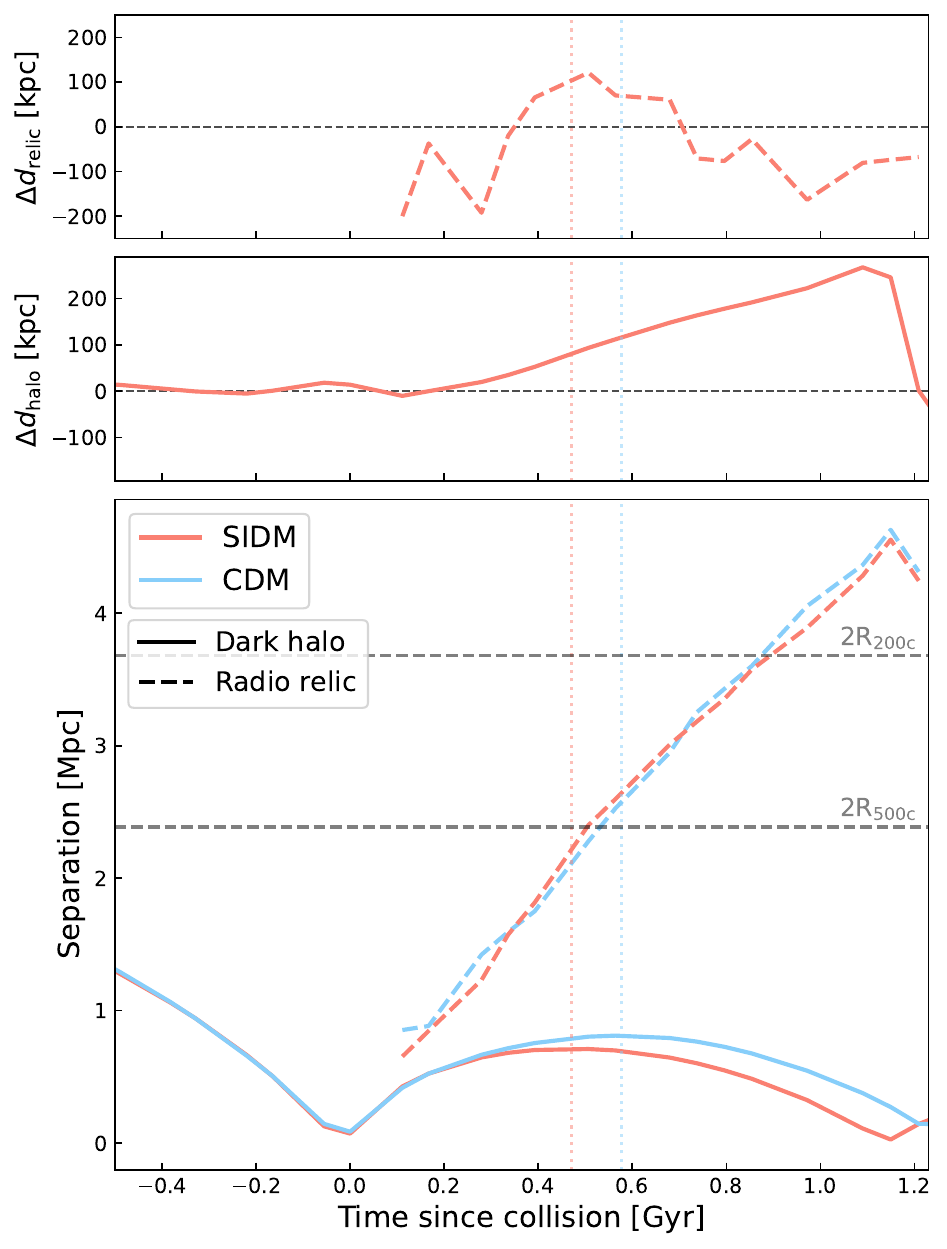}
 \caption{
 Evolution of the halo-to-halo (solid lines) and relic-to-relic (dashed lines) separation in the massive cluster merger, for $\sics=1\rm~cm^2~g^{-1}$ (\texttt{M9S1}; red), and $\sics=0\rm~cm^2~g^{-1}$ (\texttt{M9S0}; blue). The dotted vertical line marks the time of first apocenter passage in each run. The relic separation is measured for as long as both relics exceed $P_{\rm1.4GHz}>10^{22}\rm~W~Hz^{-1}$. The middle panel presents the halo separation difference between the \texttt{M9S1} and \texttt{M9S0} runs and the top panel presents the relic-to-relic separation difference between the two runs for $\rm TSC>0.2\rm~Gyr$. Overall, self-interaction changes the separation between the dark halos, while the relic separation evolves consistently regardless of $\sics$.
  }
 \label{fig:DH-relic}
 \end{figure}

\begin{figure}
 \centering
 \includegraphics[width=\columnwidth]{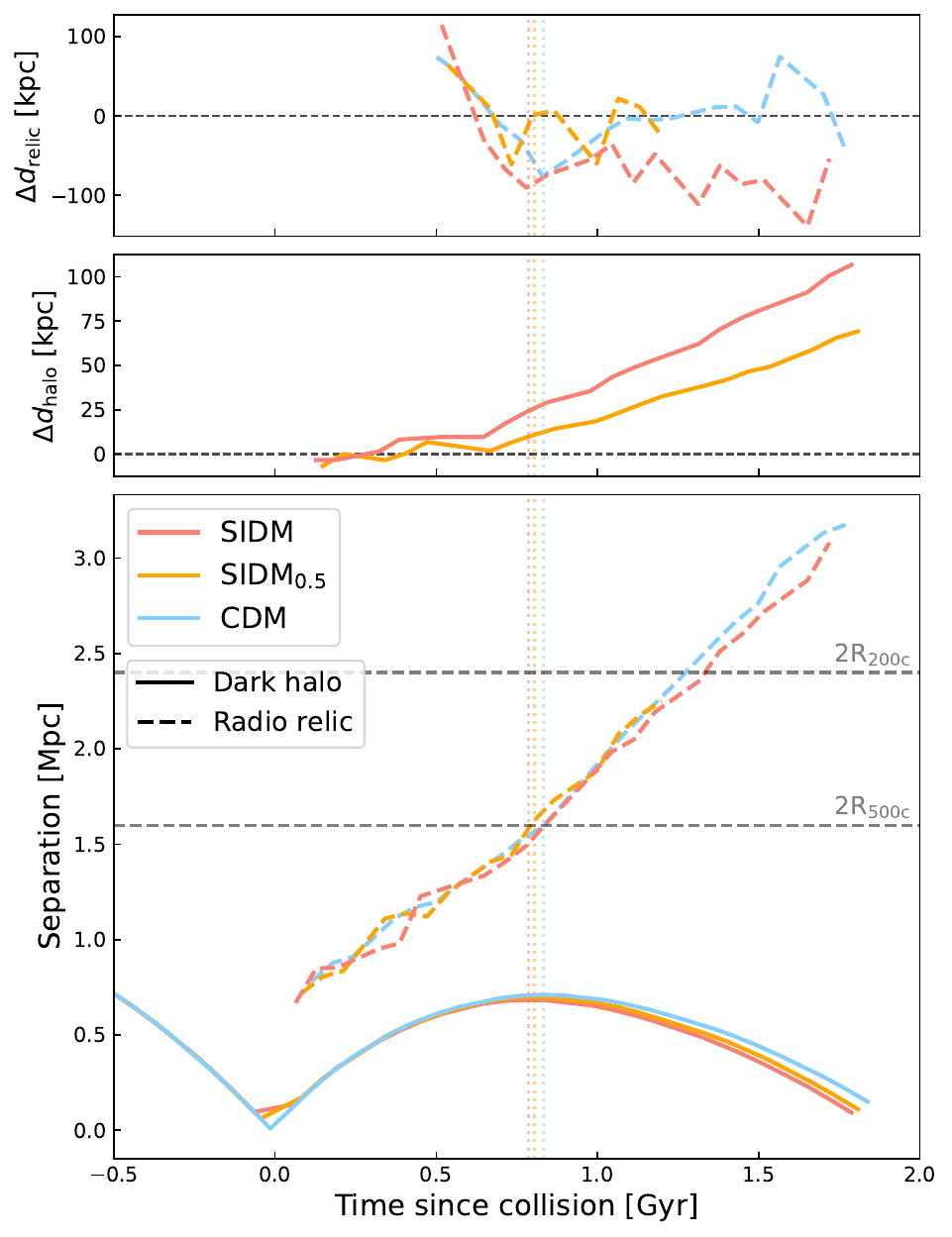}
 \caption{
 Same as Figure~\ref{fig:DH-relic} for the low-mass cluster merger with $\sics=1\rm~cm^2~g^{-1}$ (\texttt{M2S1}; red), $\sics=0.5\rm~cm^2~g^{-1}$ (\texttt{M2S05}; orange), and $\sics=0\rm~cm^2~g^{-1}$ (\texttt{M2S0}; blue). The low-mass cluster merger presents the same trend as the massive cluster merger, while the halo separation difference between the SIDM and the CDM run is smaller than in the massive cluster mergers.
  }
 \label{fig:DH-relic_low}
 \end{figure}

\subsection{Radio relic separation and halo separation}
\label{sec:shock_halo_tracer}
Figure~\ref{fig:DH-relic} shows the time evolution of the halo-to-halo and relic-to-relic separations as a function of time since the first pericenter passage, both measured from the projected maps. Self-interaction clearly presents a different impact on the two separations. Before and shortly after the pericenter passage ($\rm TSC<0.2\rm~Gyr$), the difference in halo separation with and without self-interaction ($\Delta d_{\rm halo}$) is negligible. As the merger progresses, $\Delta d_{\rm halo}$ starts to grow with time, reaching $\mytilde270\rm~kpc$ at the second pericenter passage of the \texttt{M9S1} run, before $\Delta d_{\rm halo}$ decreases as the clusters in \texttt{M9S1} enter their second outgoing phase. The first apocenter passage occurs $\mytilde0.1\rm~Gyr$ earlier in the \texttt{M9S1} run than in the \texttt{M9S0} run. 

In contrast, the relic separation evolves consistently across the dark matter models. Once the merger shocks are generated near the cluster center, they propagate nearly linearly with time. We hint at a shallower slope at the later phase, with the double relic separation increasing at the speed of $\mytilde3600\rm~km~s^{-1} ~(4300\rm~km~s^{-1})$ in $[0.1,0.5]\rm~Gyr$, and at $\mytilde3200\rm~km~s^{-1} ~(3000\rm~km~s^{-1})$ in $[0.7,1.1]\rm~Gyr$ in the \texttt{M9S0} (\texttt{M9S1}) run, which are comparable or larger than the collision velocity ($\mytilde3000\rm~km~s^{-1}$). When compared between the CDM and SIDM runs, the relic separation in the \texttt{M9S1} run expands $\mytilde20\%$ faster in the early phase and $\mytilde6\%$ slower in the later phase. At $\rm TSC\sim1.2\rm~Gyr$, the relic separation suddenly decreases as the merger shock encounters an infalling subhalo, which amplifies the radio luminosity in a localized region behind the shock front. Despite this variation in slope and subhalo encounter, the difference in relic-to-relic separation between the \texttt{M9S0} and \texttt{M9S1} runs stays within $\mytilde200\rm~kpc$, even when the relic separation reaches beyond $2R_{\rm200c}$ ($\mytilde4\rm~Mpc$). We note that the trend differs during the minor merger, as we can only identify a single relic and the halo separation evolves differently (see \textsection\ref{app:minor} for more detail).

Figure~\ref{fig:DH-relic_low} presents the same comparison for the low-mass cluster merger runs. Similar to the massive cluster merger, the halo separation decreases in the presence of self-interaction. Moreover, these runs show that the difference in the halo separation depends on $\sics$. Apocenter passage happens $\mytilde50~\rm Myr$ earlier in the \texttt{M2S1} run and $\mytilde20~\rm Myr$ earlier in the \texttt{M2S05} run, both relative to \texttt{M2S0}. When we compare the $\Delta d_{\rm halo}$ from  \texttt{M2S05} and \texttt{M2S1} at the same TSC, we find that the average ratio of the $\Delta d_{\rm halo}$ in $\rm TSC>0.7\rm~Gyr$ 
is $\Delta d_{\rm halo,\mathtt{M2S1}}/\Delta d_{\rm halo,\mathtt{M2S05}}\sim1.8$, similar to the ratio of the modeled $\sics$. At the late phase close to the second pericenter passage, $\Delta d_{\rm halo}$ reaches $\mytilde110~\rm kpc$ and $\mytilde70~\rm kpc$ when compared to the \texttt{M2S0} run with the \texttt{M2S1} and \texttt{M2S05} runs, respectively, which are smaller than the difference measured in the massive cluster merger. 

The double relic separation also evolves consistently, independent of $\sics$ in the low-mass cluster merger. After $\mytilde0.5~\rm Gyr$, once the halo begins to decelerate and merger shocks are launched, the relic separation is well described by a linear relation with a slope of $\mytilde1500~\rm km~s^{-1}$ in all runs. Although the relics in the \texttt{M2S05} run fade earlier, their separation follows the same trend until they fall below our luminosity threshold. We compare the relic separations from the simulations with the linear relation fitted to the \texttt{M2S0} relic separation over $[0.5,1.7]~\rm Gyr$. As in the massive cluster merger, the difference in relic separation across dark matter models is small. The \texttt{M2S1} run shows a relic separation $\mytilde100\rm~kpc$ smaller than this relation, but this deviation does not grow with TSC, and so its fractional contribution becomes small, especially at the late merger phase. 

We interpret the evolution of the relic separation, independent of the dark matter model, to reflect the merger shock dynamics. As discussed in \citet{2026arXiv260500093J}, merger shocks that generate radio relics are launched near the first pericenter passage, before self-interaction has significantly decelerated the halos, and their subsequent propagation is therefore largely independent of the dark matter physics. Moreover, the key merger properties that determine the shock position, including the collision velocity and total mass of the system, are primarily determined by the mass growth of the cluster inside the large-scale structures, which is unaffected by self-interaction. As a result, although stochastic variation driven by self-interactions may modify the luminosity and morphology of the radio relic, the relic-to-relic separation at the given time since the pericenter passage remains largely unaffected by the self-interaction. 

\begin{figure}
 \centering
 \includegraphics[width=\columnwidth]{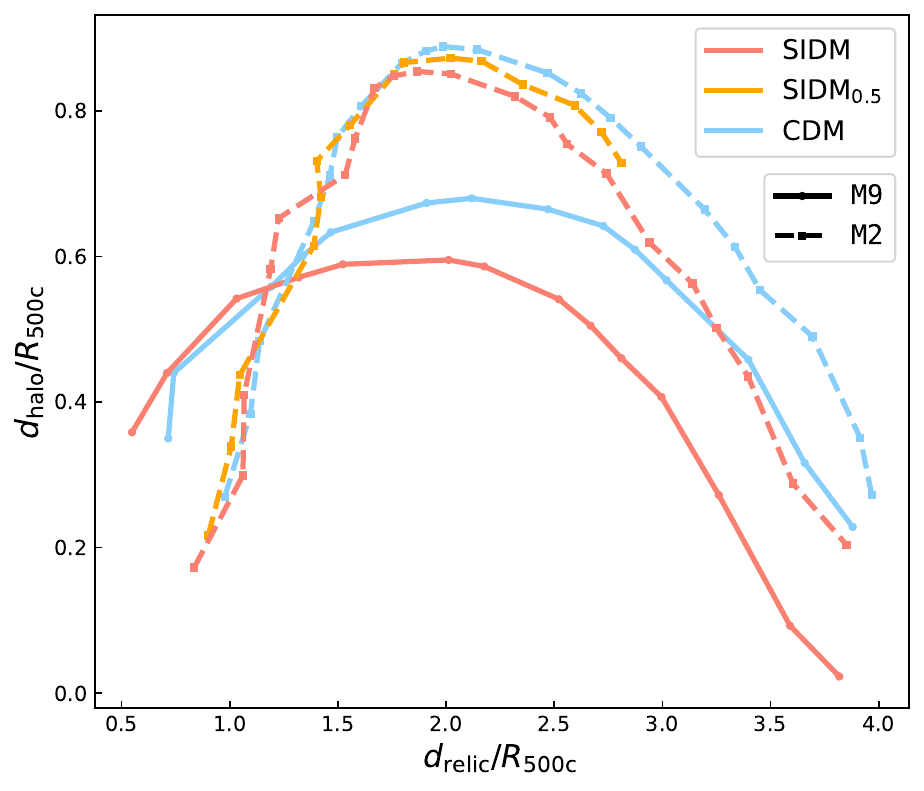}
 \caption{
 Halo-to-halo separation as a function of the relic-to-relic separation for the massive (solid lines) and the low-mass (dashed lines) cluster merger runs. All separations are normalized by $R_{\rm 500c}$ of the cluster. Blue, red, and orange lines present the CDM, $\sigma_{\rm T}/m_\chi = 1~{\rm cm^2~g^{-1}}$, and $\sigma_{\rm T}/m_\chi = 0.5~{\rm cm^2~g^{-1}}$ runs, respectively. The relic separation is presented until the relic luminosity satisfies the threshold adopted in Figures~\ref{fig:DH-relic} and~\ref{fig:DH-relic_low}. The halo separation at a fixed relic separation is smaller in the SIDM runs, and the difference increases with a larger relic separation.
  }
 \label{fig:HalovsRelic}
 \end{figure}

The evolution of double radio relic separation being independent of dark matter models can provide a reference for testing SIDM scenarios. As discussed in Figure~\ref{fig:Halo_orbits}, self-interaction modifies the halo separation during the major merger. However, halo separation itself cannot be used alone for testing the SIDM model due to its degeneracy with TSC; a halo separation smaller than the prediction from the CDM cluster mergers can be explained either with the presence of self-interaction or with a smaller (larger) TSC in its outgoing (returning) phase. As the relic separation evolves independently of self-interaction, double radio relic separations can act as a proxy of time even under the complex formation history of their host galaxy clusters. Thus, the degeneracy between the TSC and $\sics$ can be mitigated by comparing the observed halo separation with the numerical simulations at the specific double radio relic separations, implying that the double radio relic cluster mergers provide a valid testbed to test the SIDM model, as proposed by \citet[][]{2026arXiv260500093J}. 

\begin{figure*}
 \centering
\includegraphics[width=2\columnwidth]{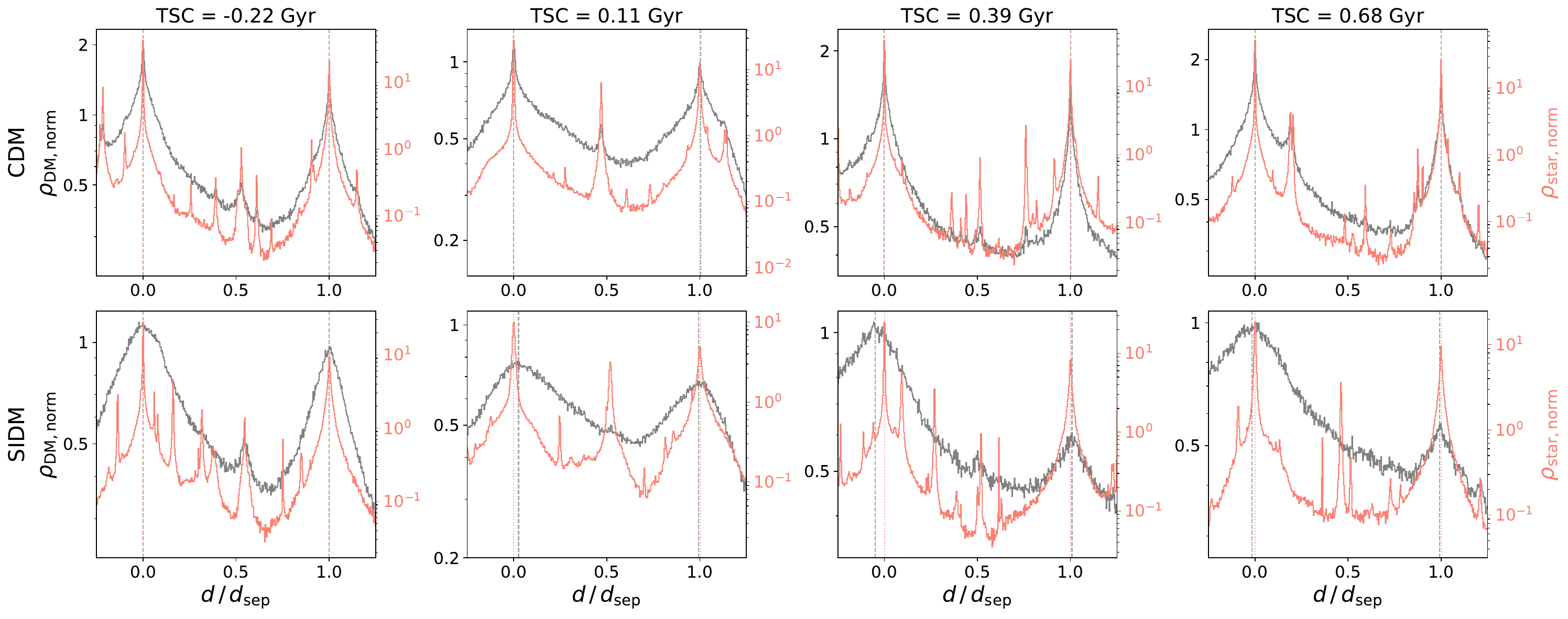}
 \caption{Normalized dark matter (grey) and stellar mass (red) distributions measured along the axis connecting the merging clusters, in \texttt{M9S0} (top) and \texttt{M9S1} (bottom). The distributions are measured within a rectangular box that samples particles within $\pm50\rm~ckpc/h$ from the axis connecting the two cluster centroids. The position is normalized by the halo separation, so the main and sub-cluster are located at 0 and 1, respectively. The dashed lines mark the position of the density peaks. The BCG of the main cluster in the \texttt{M9S1} run oscillates around the dark matter density peak, leading the dark matter at the early phase ($\rm TSC=0.11\rm~Gyr$), and lagging behind it at the later phase ($\rm TSC=0.39\rm~Gyr$). }
 \label{fig:DMstar_dist}
 \end{figure*}

Double radio relics have a practical advantage over other probes of self-interaction. Unlike probes that rely on signatures near the cluster center, double radio relics reside in the cluster outskirts, where the impact of AGN feedback is less significant \citep[but see also][]{2017MNRAS.470..240N}. Radio relics are also complementary to other probes that are most effective in the early merger phase \citep[e.g.,][but see also \textsection\ref{sec:offset}]{2014MNRAS.437.2865K}, as radio relic-based analysis would be applicable in late merger phase samples. 

Figure~\ref{fig:HalovsRelic} demonstrates how this can be applied to observations of merging clusters by presenting the halo separation as a function of the relic separation, both of which are observables.\footnote{Although the relics are identified from their three-dimensional distribution in the simulations, the relic separation can be measured from the projected map when the two relics are spatially well separated, as in a plane-of-sky merger (e.g., Figure~\ref{fig:mw_relic_zoo}).} In the early phase, the difference in halo separation between the CDM and SIDM runs is small, and this phase can be identified by the small relic separation. In the late phase, the observed halo separation can be compared with the simulated cluster mergers at the same large relic separation, where the large difference between the CDM and SIDM runs could provide a testbed for SIDM. The relic separation may differ by $\mytilde100$--$200\rm~kpc$ between the runs, as shown in Figures~\ref{fig:DH-relic} and \ref{fig:DH-relic_low}. Although this is comparable to the maximum $\Delta d_{\rm halo}$, the impact of this difference on the halo separation prediction is expected to be small, as the halo separation evolves more slowly than the relic separation. For instance, in the massive cluster merger, a relic separation difference of $\mytilde200\rm~kpc$ changes the expected halo separation by only $\mytilde4$ and $\mytilde35\rm~kpc$ at $d_{\rm relic} = 2$ and $3R_{\rm 500c}$, respectively. The $\Delta d_{\rm halo}$ induced by self-interaction at these epochs is $\Delta d_{\rm halo}\sim100$ and $\sim200\rm~kpc$, which exceeds both this shift and the typical uncertainty of $\mytilde100\rm~kpc$ in the halo separation from ground-based observations with $\mytilde30\rm~galaxies~arcmin^{-2}$ \citep[][]{2025ApJS..277...28F}. The $\Delta d_{\rm halo}$ at the second pericenter passage in the \texttt{M9} runs is much larger than these uncertainties. As the halo separation uncertainty can be reduced with deep ground-based or space-based observations \citep[e.g.,][]{2025ApJ...987L..15C}, double radio relic cluster mergers may also reveal the difference in halo separation generated by a smaller $\sics$.

Nevertheless, several systematics should be addressed when utilizing the radio relics. First of all, relic propagation depends on merger parameters such as the collision velocity and impact parameter, which can bias the TSC estimate if they are not correctly addressed. Moreover, as the halo separation evolves differently in minor mergers, the $\sics$ constraints would be biased by including the minor merger systems with double radio relics. There is also a systematic from the projection effect, as a larger viewing angle reduces the halo separation in the same direction as a larger $\sics$ (\textsection\ref{sec: projection}). Finally, the $\Delta d_{\rm halo}$ does not increase directly after the pericenter passage and shows negligible difference between the CDM and SIDM runs in the early merger phase. Combined with the halo separation that modifies after the second pericenter passage (Figure~\ref{fig:Halo_orbits}), this indicates that the relic-to-relic separation can be applied to merging clusters within a specific time window. 

Multi-wavelength analysis and numerical simulations can help mitigate this bias. As demonstrated by \citet{2026arXiv260500093J}, the impact of merger parameters can be explored using a suite of cluster merger simulations that span a range of these parameters, marginalizing the likelihood to jointly fit the merger parameters and $\sics$. Weak-lensing analysis from optical band/IR imaging can effectively classify minor merger systems \citep[e.g.,][]{2017ApJ...851...46F,2023ApJ...945...71L,2026ApJ..1005...28C}. Radio relics are rare at the early merger phase \citep[e.g.,][]{2012MNRAS.421.1868V}. Moreover, as the $\Delta d_{\rm halo}$ starts to increase when the radio relic separation deviates from the halo separation curve, such that the radio relics detach from the cluster and enter the runaway phase \citep[e.g.,][]{2019MNRAS.488.5259Z}, early-phase cluster mergers with a negligible $\Delta d_{\rm halo}$ can be excluded when we select the targets using the presence and location of radio relics relative to the halo position. The next-generation radio surveys, including SKA and LOFAR2.0, will detect faint radio relics at the late merger phase, which may include systems after the second pericenter passage. Thus, multi-wavelength observations and supporting numerical simulations that can reconstruct the merger history for accurate sample selection are essential for testing the self-interacting nature of dark matter using radio relic cluster mergers \citep[e.g.,][]{2025ApJ...984...26A}.

\begin{figure}
 \centering
 \includegraphics[width=0.85\columnwidth]{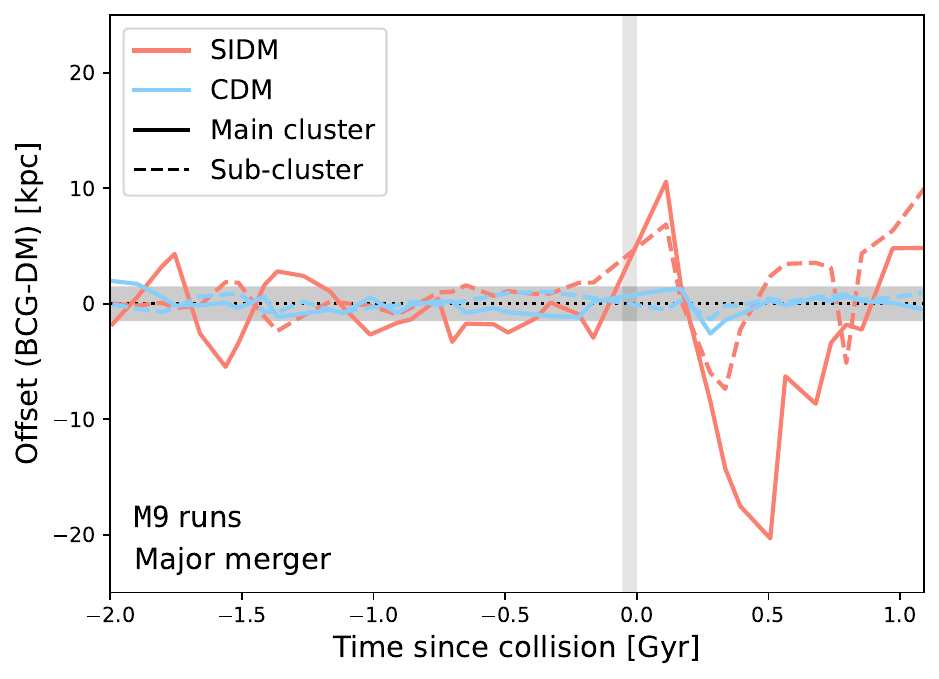}
 \includegraphics[width=0.85\columnwidth]{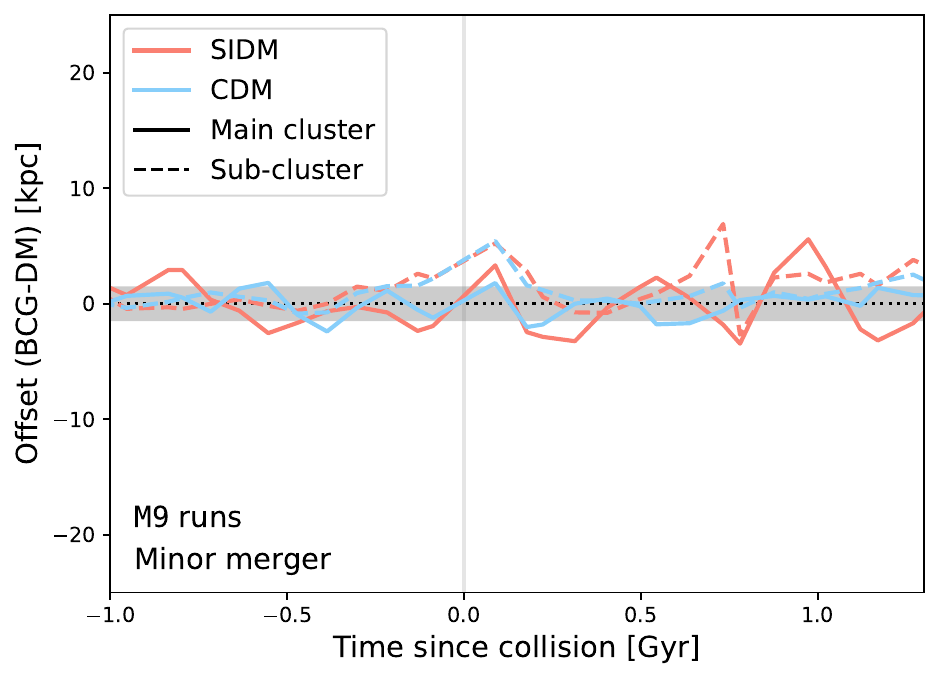}
\includegraphics[width=0.85\columnwidth]{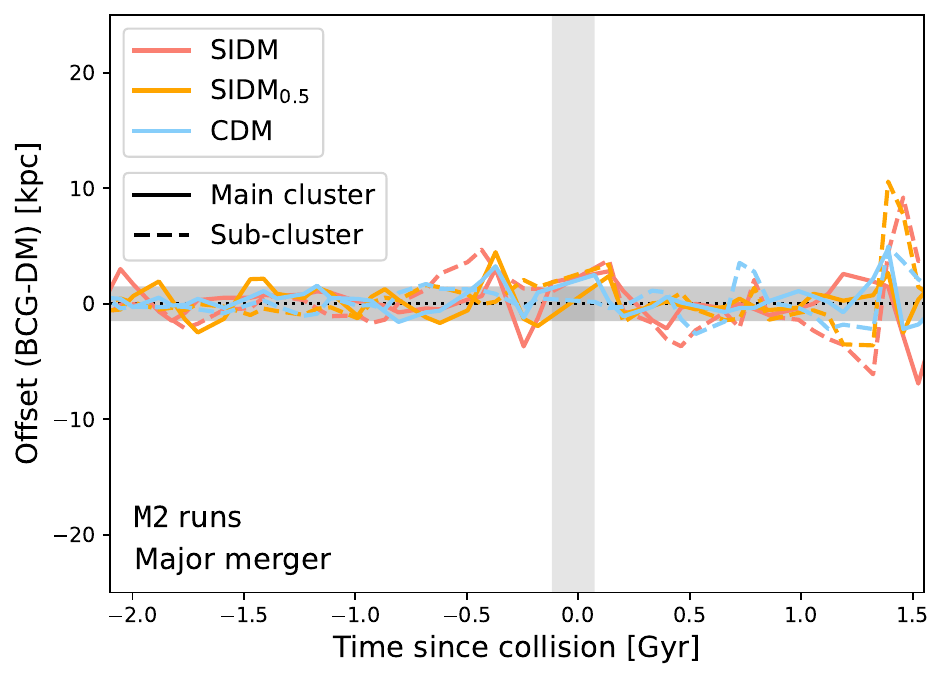}
 \caption{Spatial offset between the BCG and the dark halo centroid during the \texttt{M9} major merger (top), the \texttt{M9} minor merger (middle), and the \texttt{M2} major merger (bottom). Line color indicates the implemented dark matter model for the SIDM model with $\sics=1\rm~cm^2~g^{-1}$ (red), $\sics=0.5\rm~cm^2~g^{-1}$ (orange), and the CDM model (blue). Solid lines show the measurements from the main cluster, and dashed lines show those from the sub-cluster. The vertical grey band depicts the masked snapshot due to the small halo separation between the colliding clusters. The horizontal grey band shows the softening length of the simulation. The SIDM runs show a marginally larger BCG-dark matter halo offset than the CDM run after the pericenter passage, and the largest offset is presented in the main cluster of the massive cluster merger, likely reflecting the accumulated offset across its formation history. 
 }
 \label{fig:DMBCGoffset}
 \end{figure}

\subsection{BCG-Dark halo offset}
\label{sec:offset}
The spatial offset between the cluster galaxies and the dark matter halo center has been proposed as one of the key observables for testing the SIDM scenario using merging clusters \citep{2000PhRvL..84.3760S}. In the SIDM paradigm, the center of the galaxy distribution or the BCG is expected to lead the cluster along the collision axis, as the dark matter halos experience a drag force due to self-interaction and present a spatial offset between these two components. Multiple observations have used this observable to constrain $\sics$ \citep[e.g.,][]{2004ApJ...606..819M,2015Sci...347.1462H,2025ApJ...987L..15C}, but the reliability of the measurement has been questioned due to its degeneracy with merger history \citep{2018ApJ...869..104W}, the small spatial offset that oscillates with time \citep[e.g.,][]{2014MNRAS.437.2865K,2017MNRAS.469.1414K}, the effect of gravitational softening and systematics of the comparison between observations and simulations \citep[e.g.,][]{2024OJAp....7E..65R}. In this subsection, we track the spatial offset between the BCG and the dark halo centers in our cosmological zoom-in simulations.

Figure~\ref{fig:DMstar_dist} shows the mass distribution along the axis connecting the merging clusters of the major merger in the \texttt{M9} runs. The distance is measured along this axis and normalized by the halo separation, so 0 and 1 on the x-axis mark the position of the main and sub-cluster centers, respectively. In the CDM run, the cuspy dark matter and stellar mass peaks agree closely both before and after the collision. In the SIDM run, on the other hand, the stellar mass peaks present a small spatial offset from the smooth dark matter distribution. Before the collision, this offset is small in the main and sub-cluster. Shortly after the first pericenter passage ($\rm TSC\sim0.1\rm~Gyr$), the BCGs in both clusters lead the dark matter halo along the collision axis. By $\rm TSC\sim0.4\rm~Gyr$, the trend inverts, and the BCGs fall behind the dark matter mass peak towards the center of the system. Interestingly, our simulation presents a larger spatial offset in the main cluster, so the BCG of the main cluster at $\rm TSC\sim0.7\rm~Gyr$ is still located closer to the system center, while the BCG of the sub-cluster passes the dark halo center and is again located towards the cluster outskirts along the axis connecting the clusters.

We further measure the time evolution of the spatial offset between the BCG and the dark matter halo using projected mass maps. We first generate projected mass maps of dark matter and stellar particles in $200~\rm ckpc/h$ width with a bin size of $0.25~\rm ckpc/h$. Here, we only use dark matter and stellar particles bound to the merging clusters to avoid miscentering caused by the projected mass peaks of interlopers, yet we note that a non-negligible contribution may exist from disrupted subhalos or from the companion cluster at small separation. Starting from the center of the map, which is located at the subhalo center of each cluster, we first compute the center of mass using particles within $50\rm~kpc$ in the projection map. After shifting the center to the center of mass, we repeat the process while reducing the radius by $1\%$ for every iteration step. We stop the process when the difference between the centers from two consecutive iterations is smaller than $0.1~\rm kpc$ or when the radius reaches $10\rm~kpc$. This process typically converges at a radius of $\mytilde40~\rm kpc$ for both the CDM and SIDM runs. We confirm that the offset and its time evolution remain consistent for different choices of the initial aperture radius or with additional smoothing in the mass maps. We skip snapshots where the clusters are closer than $200\rm~kpc$ to avoid including mass from the companion cluster. 

Figure~\ref{fig:DMBCGoffset} presents the time evolution of the spatial separation between the BCG and the dark matter halo centroid during the massive and low-mass cluster mergers. We measure the distance along the axis connecting the halo centroid to the center of its companion, so that a positive (negative) value indicates that the BCG is leading (lagging) the dark matter halo along the collision axis.\footnote{We note that the first apocenter passage occurs at $\mytilde0.5\rm~Gyr$ for the major merger in the \texttt{M9} runs, and so the BCG with the negative spatial offset at late phase is leading the dark matter halo motion.} As discussed in Figure~\ref{fig:DMstar_dist}, the time evolution of the spatial offset presents the oscillatory motion of the BCGs around the dark matter halos in the SIDM mergers. In the major merger in the \texttt{M9} run, the BCG leads the dark halo by $\mytilde10\rm~kpc$ in the early merger phase ($\rm TSC\sim0.1\rm~Gyr$), then lags and reaches a larger offset of $\mytilde-20\rm~kpc$ at $\rm TSC\sim0.5\rm~Gyr$. The BCG and dark matter halo of the sub-cluster also present an oscillatory behavior, but the amplitude of the spatial offset is smaller than that of the main cluster. The median of the absolute spatial offset after the first pericenter passage and before the second pericenter passage ($\rm TSC=[0,1.2]~Gyr$) is $\mytilde11\pm5.5\rm~kpc$ and $\mytilde4.8\pm3.8\rm~kpc$ for the main and the sub-cluster, respectively. These are significantly larger than the median spatial offset of $\mytilde1.4\pm1.9\rm~kpc$ and $\mytilde0.8\pm0.8\rm~kpc$ for the main and the sub-cluster in the \texttt{M9S0} run, which is comparable to or smaller than the softening length $1.4\rm~kpc$. We note that the standard deviation is large because the offset oscillates, and so this reflects the variation across merger phases rather than the significance of the offset. 

\begin{figure}
 \centering
\includegraphics[width=\columnwidth]{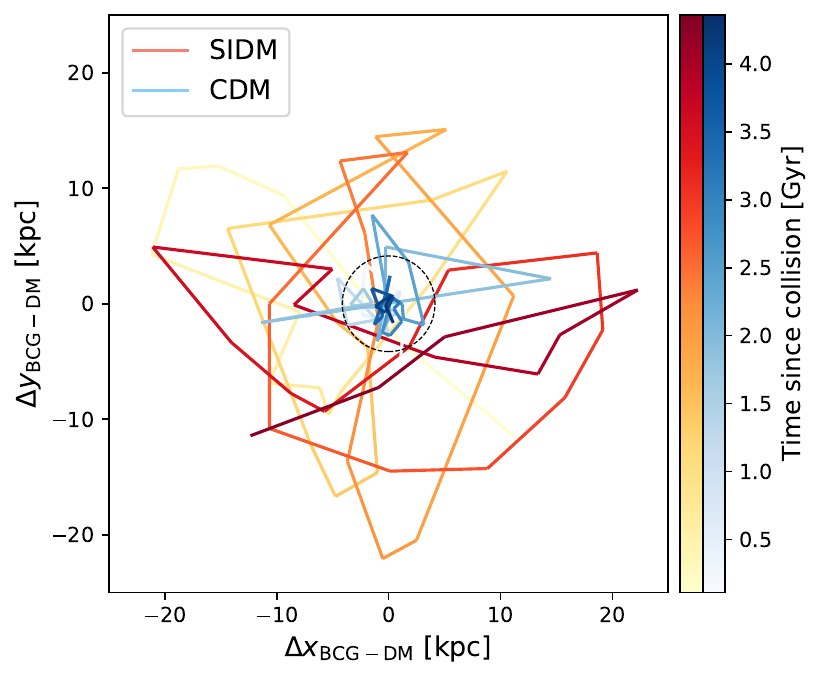}
 \caption{Long-term evolution of the location of the BCG with respect to the dark matter halo center of the main cluster in the massive cluster merger. The red and blue lines denote the SIDM run with $\sics=1\rm~cm^{2}~g^{-1}$ and the CDM run, respectively. The circle marks 2.8 times the gravitational softening length of the simulation. The BCG and the dark matter halo spatially agree in the late merger phase of the CDM run, whereas in the SIDM run the BCG continues to oscillate until $z=0$, which corresponds to $\mytilde4.4\rm~Gyr$ after the first pericenter passage.}
 \label{fig:Longer_offset}
 \end{figure}
 
The main cluster of the massive major merger presents an oscillation before the first pericenter passage. Unlike the sub-cluster, which shows a median spatial offset of $\mytilde1\rm~kpc$, the main cluster presents a median BCG-dark matter halo offset of $\mytilde3\rm~kpc$, smaller than the offset presented after the major merger, but larger than that from the \texttt{M9S0} run and the softening length $1.4\rm~kpc$. This non-zero spatial offset of the main cluster can be attributed to the minor merger that occurred $\mytilde2.5\rm~Gyr$ before the pericenter passage of the major merger. Indeed, as presented in the middle panel of Figure~\ref{fig:DMBCGoffset}, the BCG-dark matter halo spatial offset increases in the \texttt{M9S1} run after the pericenter passage of the minor merger. In the \texttt{M9S1} run, compared to the pre-merger median spatial offset of $\mytilde2.1\pm0.9\rm~kpc$ and $\mytilde0.6\pm0.9\rm~kpc$ in the main and the sub-cluster, the median spatial offset marginally increases to $\mytilde3.4\pm1.5\rm~kpc$ and $\mytilde2.7\pm2.0\rm~kpc$ for the main and the sub-cluster, respectively. Although the difference is not as dramatic as in the major merger, these offsets are larger than both the softening length and the spatial offset in the \texttt{M9S0} run, where the median spatial offset of $\mytilde1.4~(1.1)\rm~kpc$ in the main (sub-) cluster does not change with the minor merger.

The bottom panel of Figure~\ref{fig:DMBCGoffset} shows the spatial offset measured in the low-mass cluster merger of \texttt{M2} runs. As in the massive cluster merger, both SIDM runs show a spatial offset between the dark matter halo and the BCG, though their difference with the CDM runs is marginal. Excluding the late merger phase when the sub-cluster approaches to the main cluster, the median absolute spatial offset of the main and sub-clusters over $\rm TSC=[0.3,1.4]\rm~Gyr$ is $\mytilde1.1\pm1.5\rm~kpc$, $\mytilde1.2\pm1.7\rm~kpc$, and $\mytilde2.5\pm2.1\rm~kpc$ for the \texttt{M2S0}, \texttt{M2S05}, and \texttt{M2S1} runs, respectively. These offsets are larger than those measured in the pre-merger phase ($\rm TSC=[-2.1,-0.2]\rm~Gyr$), which show $\mytilde0.7\pm0.7\rm~kpc$, $\mytilde1.4\pm0.8\rm~kpc$, and $\mytilde1.3\pm1.1\rm~kpc$ for the same three runs. This increase after the first pericenter passage supports the interpretation that the implemented self-interaction creates a spatial offset between the dark matter halo and the BCG. Nonetheless, because this spatial offset is comparable to the softening length and does not clearly depend on the $\sics$, the BCG-dark matter halo offset in low-mass cluster mergers can be considered insignificant at the given resolution.

It is noteworthy that the largest spatial offset between the BCG and the dark matter halo is presented in the main cluster of the massive cluster merger. The BCG of the main cluster is marginally offset from the dark matter halo center already before the major merger, and the offset increases after the minor merger, which suggests that the mergers in its formation history may contribute to the offset. The main cluster in the \texttt{M9} runs has experienced more frequent mergers and active mass accretion than its sub-clusters or the clusters in the \texttt{M2} runs due to its large mass, similar to how the accretion history shapes the low concentration of massive clusters \citep[e.g.,][]{2008MNRAS.390L..64D}. The offset may then be further amplified by the major merger, as in Figure~\ref{fig:DMBCGoffset}. Although a larger sample of SIDM cluster mergers is needed to separate this effect from the orbital configuration of each merger, our results hint that the offset can accumulate with the merger history. We note that the magnitude of the offset may also vary with the stochasticity of galaxy formation, as suggested by the radio relic of the \texttt{M2S05} run (Figure~\ref{fig:mw_relic_zoo_low}), and a single realization per model in our simulations cannot quantify this variation.

To better understand the long-term evolution of the spatial offset, Figure~\ref{fig:Longer_offset} tracks the position of the BCG relative to the dark matter halo centroid of the main cluster after the first pericenter passage of the major merger until $z=0$. Similar to Figure~\ref{fig:DMBCGoffset}, we use particles bound to the main cluster. Across all snapshots, the BCG and dark matter halo show no significant offset in the \texttt{M9S0} run. A $\mytilde10\rm~kpc$-size offset is presented at $\rm TSC\sim2\rm~Gyr$, which can be attributed to the particles bound to the sub-cluster that start to merge into the main cluster (see the halo separation in Figure~\ref{fig:Halo_orbits}). In contrast, the BCG oscillates with respect to the main cluster center for a long timescale in the \texttt{M9S1} run. At $z=0$, when the time since the first pericenter passage reaches $\mytilde4.4\rm~Gyr$, the BCG still presents a $\mytilde20\rm~kpc$ offset from the dark matter halo, which is comparable to the spatial offset at the early merger phase. This demonstrates that the spatial offset between the dark matter halo and the BCG does not damp effectively with time, and so the offset may accumulate with the growth of cluster mass. 

Our simulation supports the previous studies that predict long-lasting wobbling of BCGs in the SIDM clusters \citep[e.g.,][]{2017MNRAS.469.1414K}. Moreover, a larger spatial offset in more massive clusters has been reported from the statistical analysis of galaxy clusters \citep[e.g.,][]{2019MNRAS.488.1572H}. As demonstrated by these works, the statistical analysis of the spatial offset in clusters may provide a stringent testbed for SIDM. Nonetheless, our simulation presents a $\mytilde3\rm~kpc$ offset in the main cluster before the major merger, when its mass was $M_{\rm 200c}\sim5\times10^{14}M_{\odot}$. This implies that the spatial offset may strongly depend on its mass growth history, and so the statistical average of massive clusters would not always present a significant offset even with a large $\sics$ as $1\rm~cm^2~g^{-1}$. Instead, as the large spatial offset does not damp after the major merger, we suggest statistical analysis of the BCG-dark matter halo offset in the massive clusters with a hint of old major merger activity, which can be identified via X-ray kinematics \citep[e.g.,][]{2024ApJ...974..234B,2026A&A...707A.109Z,2026ApJ...996..106W} or mass distribution \citep[e.g.,][]{2025NatAs...9..925H}, would better reveal the offset if it exists.

Similar to the radio relic probes, applying the BCG--dark matter halo offset to observations requires addressing several systematics. The large offset in the massive merger remains detectable in the smoothed total mass distribution of gravitational lensing analysis (\textsection\ref{app:BCGTotal}), but interlopers along the line of sight can bias the measured offset, since observations cannot preselect the gravitationally bound particles as in our analysis. In addition, whether the offset remains detectable when averaged over mergers with different configurations and phases cannot be established from our two systems. A larger suite of simulated cluster mergers is therefore needed to assess the BCG--dark matter halo offset as an SIDM probe.

\begin{figure}
 \centering
\includegraphics[width=\columnwidth]{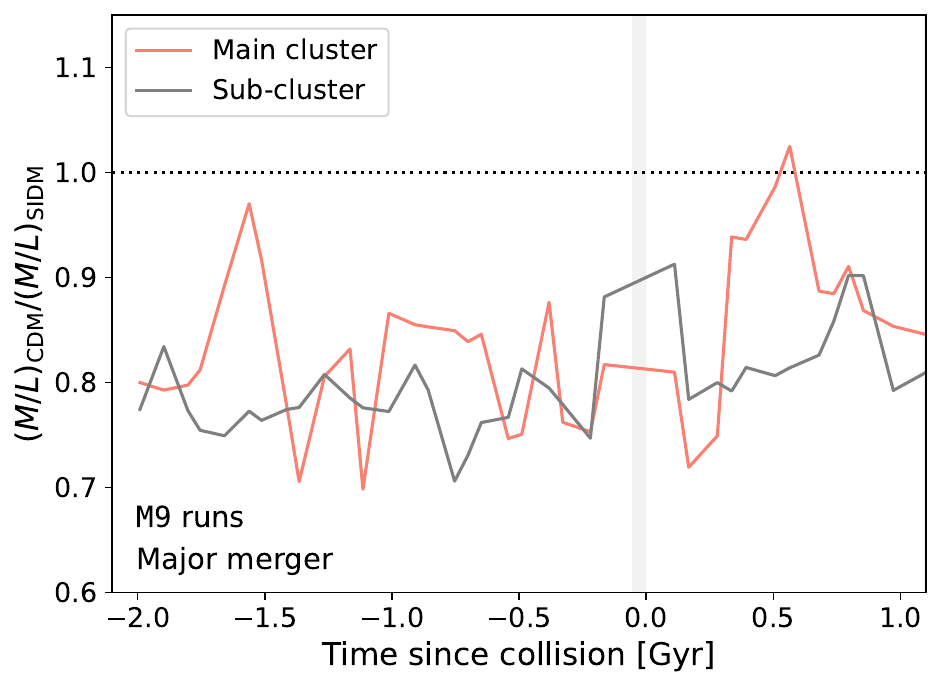}
\includegraphics[width=\columnwidth]{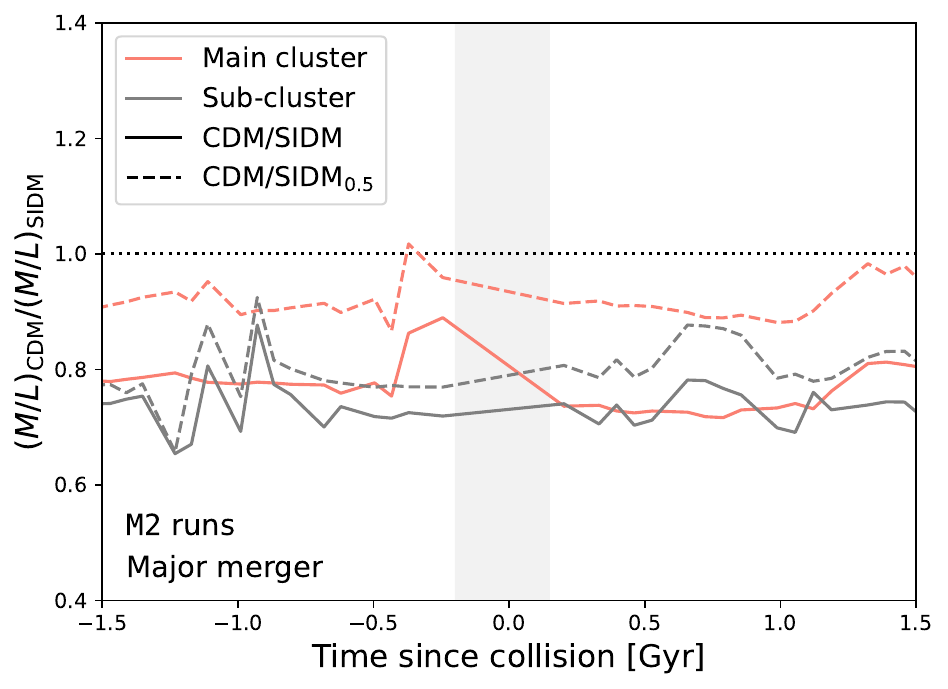}
 \caption{Ratio of the M/L between the CDM and SIDM runs across the major merger events. The shaded region marks the epoch when the halo separation is smaller than $300~\rm kpc$. The ratio of the M/L is converted from the ratio of the dark-matter-to-stellar mass, as the two simulations present a similar stellar population. Although the mass ratio differs with self-interaction, this impact does not evolve systematically with merger phase.
 }
 \label{fig:ML_ratio}
 \end{figure}

\subsection {Mass-to-light ratio}
\label{sec:mlratio}

The M/L of a cluster is expected to change with self-interaction \citep[][]{2008ApJ...679.1173R}. Similar to the BCG-dark halo offset, previous studies predict that the spatial offset between the collisionless galaxies and the self-interacting dark matter halo increases the M/L when measured from the dark halo center. As our simulation implements galaxy formation models self-consistently with cluster formation, we estimate the time evolution of the cluster M/L before and after the merger.

To analyze the impact of self-interaction on the M/L, we measure the dark matter and stellar mass in the projection map within a fixed radius of $150\rm~kpc$ from the cluster center. This dark-matter-to-stellar-mass ratio can be converted to M/L using the mass-to-light ratio of the stellar population. We find that the median stellar age agrees between the CDM and SIDM runs to within $\mytilde0.2\rm~Gyr$ in both \texttt{M9} and \texttt{M2} runs, with an average age at $z=0$ of $\mytilde11\rm~Gyr$. We therefore assume the same mass-to-light ratio applies to the stellar populations in both runs, so the difference between the mass ratios of the CDM and SIDM runs, $(M_{\rm DM}/M_{\rm star})_{\rm CDM}/(M_{\rm DM}/M_{\rm star})_{\rm SIDM}$, can be interpreted as the difference in their M/L. We note that assuming a stellar population mass-to-light ratio of $\mytilde3$, the M/L within $R_{\rm 200c}$ at $z=0$ is $\mytilde190~(\mytilde220)$ in the \texttt{M9S0} (\texttt{M9S1}) run. 

Figure~\ref{fig:ML_ratio} presents the time evolution of the M/L difference between the CDM and SIDM runs. We find that the M/L is generally smaller in the CDM simulations than in the SIDM runs. In both the massive and the low-mass cluster merger runs, the median ratio of $(M/L)_{\rm CDM}$ to $(M/L)_{\rm SIDM}$ ranges within [0.77, 0.87]. This trend is weaker in the \texttt{M2S05} run, where the median ratio of the main cluster is $\sim$0.92, closer to unity. The larger M/L ratio in the SIDM run is mainly driven by the larger stellar mass in the CDM runs. Within the $150~\rm kpc$ radius, the enclosed dark matter mass in the CDM and the SIDM run agrees within $\mytilde5\%$ before the collision, while the stellar mass in the CDM run is $\mytilde20\%$ higher than that from the SIDM run. 
After the collision, the stellar mass in the SIDM run stays consistently lower than the CDM run, while the dark matter mass is reduced within the fixed radius in the SIDM run by the redistribution from scattering. As a result, the M/L of the SIDM run becomes comparable to that of the CDM run at $\rm TSC\sim0.5\rm~Gyr$ in the main cluster of the \texttt{M9} run. A similar trend can also be identified in the sub-cluster of the low-mass cluster merger at $\rm TSC\sim0.7\rm~Gyr$ in both \texttt{M2S1} and \texttt{M2S05} runs. However, this trend is unclear in the sub-cluster of the major merger, or is presented at a later epoch in the main cluster of the minor merger (TSC$\sim1.3\rm~Gyr$). 

Our results show that the M/L cannot provide a robust reference for the SIDM. Although the M/L is higher in the SIDM runs and the difference tends to decrease after the pericenter passage, the trend is not consistent between the main and the sub-cluster. Moreover, the impact of self-interaction on M/L is also unclear in the \texttt{M2S05} run. We can also identify a similar level of fluctuation at the pre-merger phase. For instance, the M/L of \texttt{M9S1} becomes comparable to that of \texttt{M9S0} at $\rm TSC\sim -1.6\rm~Gyr$, possibly driven by a subhalo that resides close to the main cluster center. Given this unclear trend and the degeneracy with the interlopers, the M/L would not be an ideal probe for testing SIDM in merging clusters.

\section{Discussion}
\label{sec: discussion}
\subsection{Resolution dependence}
\label{sec: Resolution}

\begin{figure}
 \centering
 \includegraphics[width=\columnwidth]{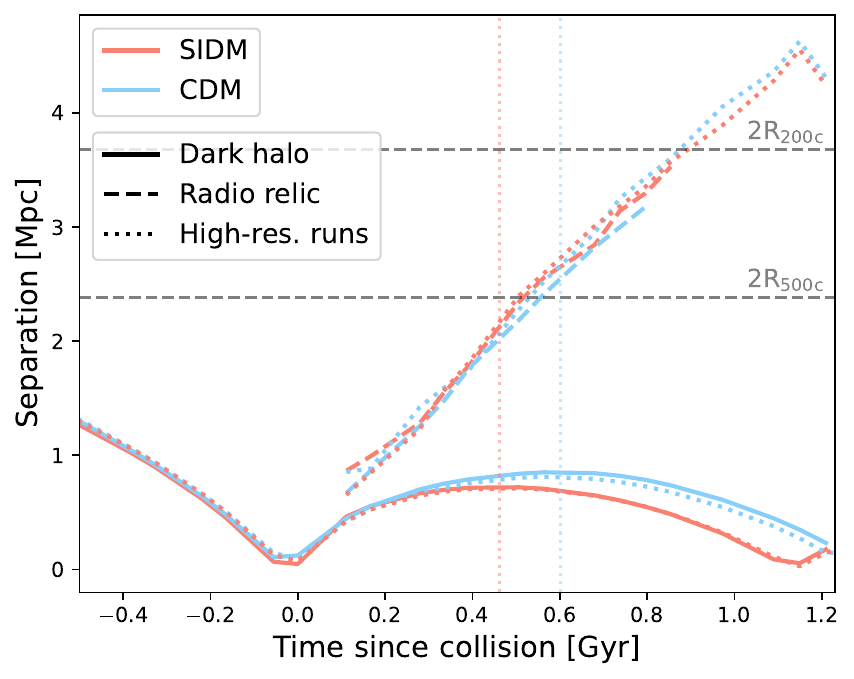}
 \includegraphics[width=\columnwidth]{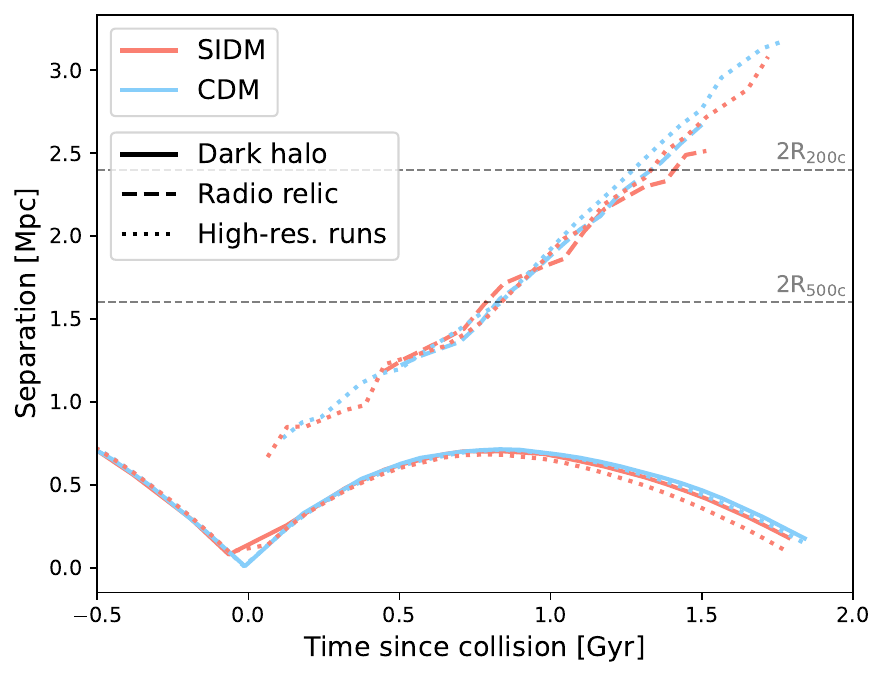}
 \caption{
 Same as Figure~\ref{fig:DH-relic}, measured in the low-resolution simulation for the massive (top) and low-mass cluster merger (bottom). The red and blue plots show the measurements from the SIDM and the CDM runs, respectively. The dotted lines annotate the halo-to-halo and relic-to-relic separation in their high-resolution counterpart. The low-resolution simulation reproduces a similar relic-to-relic and halo-to-halo separation in the massive cluster mergers, while the impact of self-interaction is underestimated in the low-resolution run of the low-mass cluster merger.
 }
 \label{fig:LowRes_relic}
 \end{figure}

In this study, we run cosmological zoom-in simulations of massive and low-mass cluster mergers with and without self-interacting dark matter. The dark matter mass resolution of $6.1\times10^{7}~M_{\odot}$ samples the clusters with a particle number on the order of $\mytilde10^7$, providing high spatial resolution that can accurately track the evolution of the cluster merger observables. In \textsection\ref{sec: SIDM_tracers}, we find that radio relic separations or BCG-dark matter halo offsets can act as SIDM probes even under the complex formation history of clusters inside large-scale structures. However, as the self-interaction is boosted during the pericenter passage, the SIDM signatures may depend on numerical resolution. To assess the robustness of the observational probes discussed in this work, we repeat the analysis using the low-resolution runs, which have the mass resolution of $m_{\rm DM}\sim5\times10^{8}~M_{\odot}$ and $m_{\rm baryon}\sim9\times10^{7}~M_{\odot}$, a factor of $\mytilde8$ higher than those of the main runs. The low-resolution run uses the same initial conditions as the high-resolution runs and so presents the identical merger event. We skip the BCG-dark matter halo offset measurement of the low-mass cluster merger, as their spatial offset was already comparable with the softening length of the high-resolution runs.

Figure~\ref{fig:LowRes_relic} presents the time evolution of halo and relic separations in the low-resolution simulations. The overall trend is consistent with the high-resolution runs and support the convergence of our result. In the low-resolution run of the massive clusters (\texttt{M9S1Low} and \texttt{M9S0Low}), the halo separation in the SIDM and the CDM runs, and the double relic separations are comparable to their high-resolution counterpart. The halo separation in the CDM run is larger in the low-resolution run, while the difference is only marginal. We find that the double relic fades more quickly in the low-resolution runs. Although the relic separation follows that of the high-resolution run, the relic luminosity falls below the threshold at $\rm TSC\sim0.8\rm~Gyr$, $\mytilde0.4\rm~Gyr$ earlier than that of the high-resolution run. These consistent relic locations and different radio luminosities agree with the resolution test in \citet{2024A&A...686A..55L} using TNG300-1 and TNG300-2. 

\begin{figure}
 \centering
 \includegraphics[width=\columnwidth]{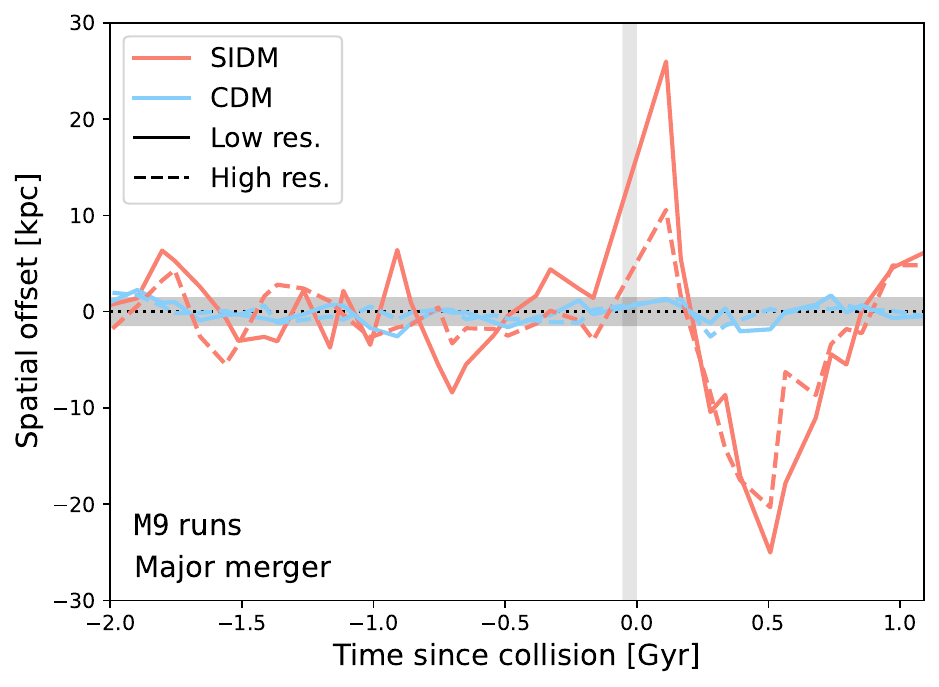}
 \caption{
 Same as Figure~\ref{fig:DMBCGoffset}, measured in the low-resolution run of the major merger event of the massive cluster. The oscillatory trend and the $\mytilde20\rm~kpc$ offset are reproduced in the low-resolution runs, supporting the robustness of the result against the numerical resolution.
 }
 \label{fig:LowRes_offset}
 \end{figure}

On the other hand, we find that the halo separation evolves differently in the low-mass cluster merger. Although the halo separation in the SIDM run is smaller than that of the CDM run, the low-resolution SIDM run (\texttt{M2S1Low}) presents a similar halo separation to that of the \texttt{M2S0} run. As a result, compared to the $\mytilde110~\rm kpc$ difference in the halo separation near the second pericenter passage, the difference in the halo separation is reduced to $\mytilde50~\rm kpc$ in the low-resolution run. For relic separation, the low-resolution run presents a marginally smaller relic separation at the same TSC, while their luminosity drops at the earlier TSC, as in the massive cluster merger runs. The weaker impact of the self-interaction in low-resolution runs of the low-mass cluster merger can be interpreted with the number of dark matter particles at the cluster centers. As discussed in Figure~\ref{fig:Rate}, the low-mass cluster merger run presents the impact of self-interaction at $\sim0.02R_{\rm 200c}$ ($\mytilde20\rm~kpc$), where the interaction occurs on the dynamical timescale. However, in the low-resolution simulations, this region is now similar to the convergence radius in the \texttt{M2S0Low} run ($\mytilde20\rm~kpc$), and so the self-interaction is modeled with an insufficient number of dark matter particles, underestimating the impact during the merger.

Figure~\ref{fig:LowRes_offset} further presents the BCG-dark matter halo spatial offset measured in the $\texttt{M9S1Low}$ and $\texttt{M9S0Low}$ runs. The overall trend is similar to that of the high-resolution run. The BCG-dark matter halo offset in the CDM run is negligible, whereas the spatial offset increases to $\mytilde20\rm~kpc$ after the pericenter passage in the SIDM runs. The low-resolution run also presents the oscillatory behavior of the BCG-dark matter halo spatial offset, where the BCG lags behind the dark matter halo shortly after the pericenter passage, which then passes the dark matter halo and leads its motion, as discussed with the high-resolution runs. 

This test shows that the overall trends of the observational probes discussed in this work are consistent between the high- and low-resolution runs of the massive cluster merger. At the lower resolution, the halo-to-halo and relic-to-relic separations follow those measured from the high-resolution run, and the oscillatory spatial offset between the BCG and dark matter halo is also recovered. On the other hand, the impact of self-interaction is underestimated in the low-resolution run of the low-mass cluster merger, where the region in which the interaction rate becomes significant is not well resolved. The difference between the massive and low-mass cluster mergers, which could partly be numerical, should therefore be interpreted with care. This caution extends beyond zoom-in simulations, highlighting the importance of checking the resolution dependence in statistical analyses of SIDM cluster mergers.

\begin{figure*}
 \centering
\includegraphics[width=2\columnwidth]{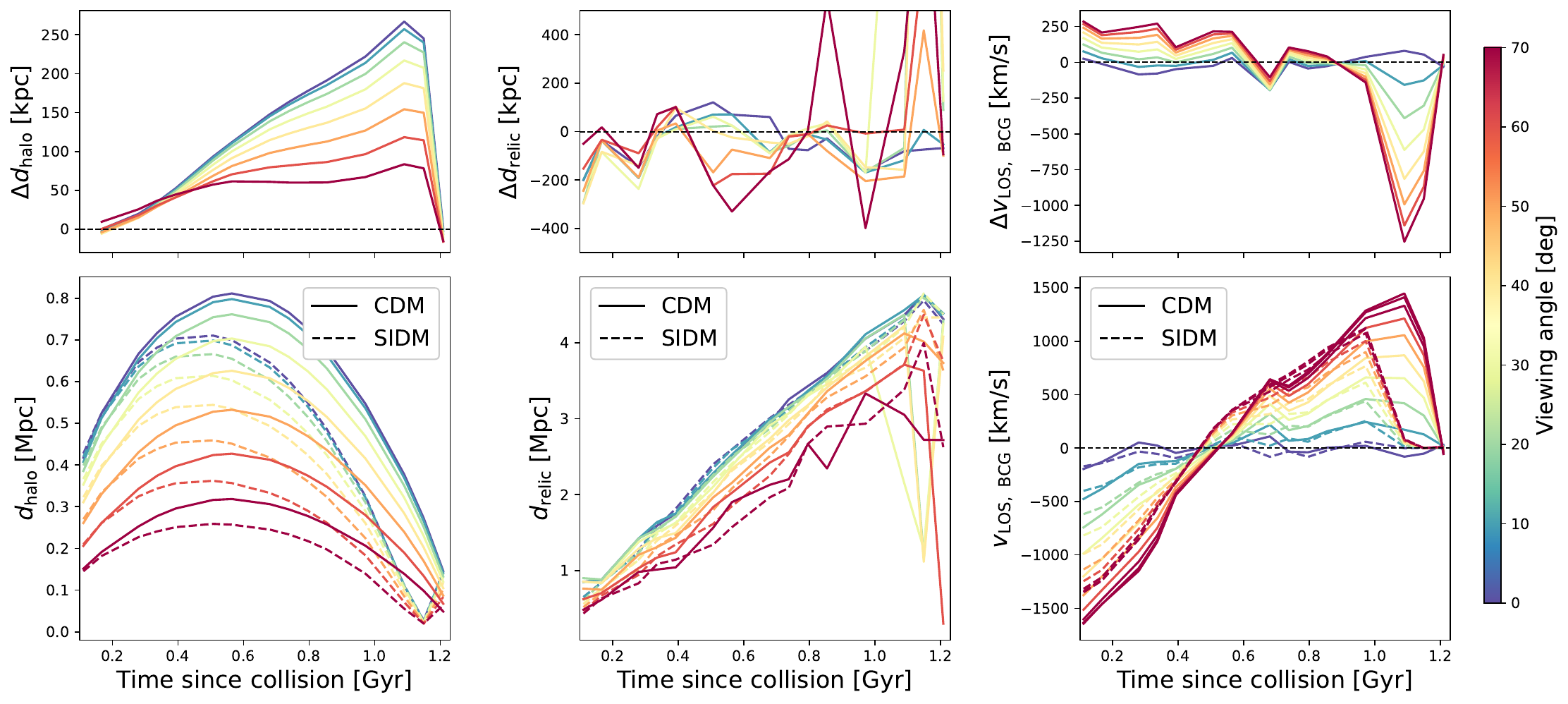}
 \caption{
Halo (left) and relic separation (middle), and relative line-of-sight velocity of the BCG (right) in the CDM (solid) and SIDM with $\sics = 1~\rm cm^{2}~g^{-1}$ (dashed) runs under different projection angles. The viewing angle is defined from the plane-of-sky merger, with a viewing angle of $90\degr$ corresponding to projection as a line-of-sight merger. The top row shows the difference between the CDM and SIDM runs, where the halo and relic separation residuals are derived as in Figure~\ref{fig:DH-relic} and the BCG velocity residual is obtained by interpolating the CDM velocities onto the SIDM time axis before subtracting. While the halo separation shows a clear viewing-angle dependence, the relic separation remains largely insensitive to projection. The relative LOS velocity of the BCG provides a complementary observable that can help mitigate the projection-induced degeneracy in the halo separation. }
 \label{fig:Projection}
 \end{figure*}

\subsection{Projection effect in the self-interaction constraint}
\label{sec: projection}

In this work, we assume the cluster mergers are viewed as plane-of-sky mergers. This could be a plausible assumption for the cluster mergers selected based on double radio relics \citep[e.g.,][]{2019ApJS..240...39G}, which, however, may include outliers with a large viewing angle \citep[e.g.,][]{2018ApJ...862..160W}. In this subsection, we test the impact of nonzero viewing angles on using the double radio relic separations for testing SIDM.

Figure~\ref{fig:Projection} summarizes the projection effect on the halo and relic separations. We note that both radio relics remain identifiable at all viewing angles, even when the double radio relics significantly overlap, as we identify the cells associated with each radio relic from their three-dimensional distribution. This classification would not be available in observations, especially when the viewing angle is large and the relic-to-relic separation is small. For this reason, we have varied the viewing angle up to 70 degrees, with a caution on the interpretation that a large viewing angle might not be identified as a double radio relic cluster merger from observations. 

We can see that the viewing angle has a different impact on the halo and relic separation. Halo separation decreases with increasing viewing angle, as the cluster merger is viewed closer to a line-of-sight merger. As a result, the first apocenter separation, and so the difference between the CDM and the SIDM run, decreases with increasing viewing angle. On the other hand, the difference in the relic separation is relatively smaller than that of the halo separation. For viewing angles smaller than 50 degrees, the relic-to-relic separation follows a similar evolution to that from the plane-of-sky merger. This weak dependence on the viewing angle can be attributed to the spherical shell-like morphology of the radio relics and the separation measured between the edges of the radio relics \citep[e.g.,][]{2026arXiv260500093J}. As a result, the difference in the relic separation between the CDM and the SIDM run presents a similar trend, even when the system is viewed with a large viewing angle. 

The viewing angle clearly introduces an additional degeneracy with the $\sics$. The viewing angle, which decreases the expected halo separation while presenting a consistent relic separation, has the same impact as assuming a larger $\sics$. As demonstrated in \citet{2026arXiv260500093J}, this degeneracy can be addressed with prior knowledge on the mass of the system. Since the halo and relic separation at a given TSC depends on the system mass, an independent mass estimate allows us to predict both separations at the given TSC under the CDM paradigm. By comparing this prediction with the observed separations, one could distinguish whether the smaller separation from observation is due to viewing angle or self-interaction. 

The second approach is to utilize the relative line-of-sight (LOS) velocity of the BCGs. Here, we measure the LOS velocity of the BCG with the mass-weighted average velocity of the stellar particles within 200 kpc of each cluster. As shown in the right column of Figure~\ref{fig:Projection}, the relative LOS velocity of the BCG is a sensitive function of the viewing angle. When the cluster is viewed as a plane-of-sky merger, by definition, both the CDM and SIDM runs present a relative LOS velocity between the BCGs close to zero. The relative velocity rapidly increases with the viewing angle due to the high collision velocity of the system. Moreover, since the initial collision history is similar in the CDM and SIDM runs, the increment in the LOS velocity with viewing angle follows a similar trend. Because this dependence on viewing angle differs from that of the halo and relic separation, the relative LOS velocity of the BCG can mitigate the degeneracy of the viewing angle.

This analysis demonstrates that multi-wavelength observations are required for breaking the degeneracy between the self-interaction cross-section and the merger geometry. We caution, however, that the sensitivity of each observable varies with the merger phase. For instance, while the relative LOS velocity of the BCG constrains the viewing angle, its signal diminishes near the first apocenter passage as the orbital velocity approaches zero. Multi-wavelength observables that evolves distinctly with TSC \citep[e.g.,][]{2024ApJ...968...74S,2025ApJ...995L..50H} or additional properties such as the polarization fraction of the radio relics \citep[e.g.,][]{2002MNRAS.331.1011E} may provide complementary constraints at these phases. Nevertheless, these indicators may not always lead to a consistent picture, as suggested in the recent multi-wavelength study of PSZ2~G181.06+48.47 \citep[][]{2025ApJ...984...26A,2025ApJ...984...24S,2025ApJ...984...25R}. More sophisticated modeling that can jointly fit multiple observables across the full merger history may therefore be necessary to fully exploit the constraining power of merging clusters on $\sics$.

\section{Summary  \& conclusion}
\label{sec: summary}

In this work, we present cosmological magnetohydrodynamical zoom-in simulations of two cluster mergers, one massive ($M_{\rm 200c}\sim9\times10^{14}~M_\odot$) and one low-mass ($M_{\rm 200c}\sim2\times10^{14}~M_\odot$) cluster mergers, performed with CDM and SIDM models assuming an isotropic, velocity-independent cross-section of $\sics=1~\rm cm^{2}~g^{-1}$, with an additional $\sics=0.5~\rm cm^{2}~g^{-1}$ run for the low-mass cluster merger. Both systems are selected from the TNG-Cluster suite, which provides optimal conditions for exploring the impact of self-interaction on cluster mergers, and are re-simulated with the same galaxy formation model and mass resolution as the parent simulation, allowing us to track the merger observables at high resolution. We investigate the impact of dark matter self-interaction on the observables of the merging clusters, with a focus on identifying reliable tracers of $\sics$ applicable to multi-wavelength observations. Our main findings are summarized as follows:
\begin{itemize}

\item Self-interaction thermalizes the dark matter halo, producing an isothermal core within $\mytilde0.08R_{\rm 200c}$ ($\mytilde150~\rm kpc$) where the expected interaction rate is significant within the dynamical timescale of the system. The number of subhalos decreases with the presence of self-interaction, and the impact is larger in the central region, where the self-interaction is more frequent. The inclusion of baryons weakens the impact of self-interaction due to the additional gravitational potential. The dark matter halo is more spherical in the SIDM case, while the CDM and SIDM halos become comparably spherical in the innermost region.

\item The impact of self-interaction on the stellar mass distribution is weaker than that on the dark matter distribution. The overall stellar mass function and ellipticity remain similar between the SIDM and the CDM runs. Nevertheless, the more active tidal disruption in the SIDM clusters can present more prominent shell-like substructures in the intracluster light distribution.

\item We confirm that the relic-to-relic separation provides a reliable, dark-matter-model-independent reference for testing the SIDM model, as proposed by \citet{2026arXiv260500093J}. The relic separation evolves consistently with time since collision in both the massive and low-mass cluster mergers, so comparing the halo separation at the same relic separation can distinguish between dark matter models.

\item The impact of self-interaction on the halo separation changes sign in minor mergers and after the second pericenter passage. We interpret this as the drag force that self-interaction exerts on colliding halos competing with the weakened dynamical friction in SIDM halos. As a result, the halo separation of minor mergers and of cluster mergers after their second pericenter passage should be interpreted with caution when testing the SIDM scenario.

\item The BCG presents a spatial offset from the dark matter halo after the major merger, reaching $\mytilde20\rm~kpc$ in the massive cluster merger. The largest offset is found in the main cluster rather than the sub-cluster, which may reflect the more frequent merger activity that the main cluster has experienced throughout its growth history.

\item The BCG--dark matter halo offset continues to oscillate beyond $4\rm~Gyr$ after the first pericenter passage in our simulated cluster. As not all massive clusters experience major mergers, statistical analysis of the offset in massive clusters with a hint of old major merger activity may reveal the signature of self-interaction most significantly.

\item The mass-to-light ratio is higher in the SIDM runs, primarily due to their lower stellar content. Although the scattering of dark matter modifies the M/L after the merger, the trend is not consistent across our sample, and a comparable level of fluctuation is already present before the merger, which limits the use of M/L as an SIDM probe.

\item The resolution test shows that the SIDM probes discussed in this work are consistent between the high- and low-resolution runs of the massive cluster merger. However, the low-resolution run of the low-mass cluster merger underestimates the impact of self-interaction, as the region where self-interaction occurs is comparable to the convergence radius of the simulation. The impact of self-interaction should therefore be interpreted with care unless the central regions are  well resolved.

\item The viewing angle introduces a degeneracy with $\sics$ when using the halo separation as an SIDM probe, since a larger viewing angle and a larger $\sics$ both reduce the apparent halo separation similarly. This degeneracy can be mitigated with multi-wavelength observations, including an independent mass estimate or the relative line-of-sight velocity of the BCGs, both of which depend on the viewing angle differently than the halo separation does. 

\end{itemize}

Our results are qualitatively consistent with previous idealized simulations. The relic separation evolves regardless of the self-interaction model, as proposed by \citet{2026arXiv260500093J}, and is comparable to their relic separation at a given TSC. The BCGs also oscillate around the dark matter halo for several Gyr after the pericenter passage \citep[e.g.,][]{2017MNRAS.469.1414K,2023MNRAS.523.5915F}, with an offset of $\mytilde20\rm~kpc$ for $\sics=1\rm~cm^2~g^{-1}$ that is comparable to the major merger in our simulations. Although the merger configurations are not identical, these similarities suggest that idealized simulations can provide reasonable predictions for the signatures of SIDM in merging clusters. However, the mass inflow from the large-scale structure and the accumulated self-interaction during the cluster formation are not captured in isolated setups. They may alter the relic separation at the late merger phase and produce a nonzero BCG offset before the first pericenter passage. Thus, caution is needed when using idealized simulations for a quantitative comparison with observed clusters.

Based on these findings, we suggest criteria for selecting cluster mergers to test the self-interacting nature of dark matter. For the halo separation, the sample should consist of major mergers identified with double radio relics, at a relic separation beyond $\mytilde2R_{\rm 500c}$ and before the second pericenter passage. For the BCG--dark matter halo offset, massive clusters with evidence of an old major merger are promising targets. As these criteria are drawn from two simulated mergers, they provide a starting point for designing SIDM merger samples. 

Our simulations further highlight the importance of capturing massive cluster mergers in cosmological simulations with SIDM. Our results show that massive clusters that underwent a major merger can exhibit unique features, such as shell-like substructures in the ICL and a BCG--dark matter halo offset. However, our sample of two cluster mergers cannot determine whether these features would also be visible in massive clusters without a major merger, or how the impact of self-interaction varies with the merger configuration. Statistics of cluster mergers with SIDM are therefore needed to interpret the signatures of self-interaction in the cluster mergers observed by upcoming surveys such as the SKA and the Rubin Observatory.

\begin{acknowledgements}
The authors thank Volker Springel and Fangzhou Jiang for valuable discussions. This work was supported by the National Research Foundation of Korea (NRF) programs: 2022R1A2C1003130, RS-2024-00340949, RS-2025-00437358, and RS-2026-25607498. DN acknowledges funding from the Deutsche Forschungsgemeinschaft (DFG) through an Emmy Noether Research Group (grant number NE 2441/1-1). This work is co-funded by the European Union (ERC, COSMIC-KEY, 101087822, PI: Pillepich).
We thank the Korea Institute of Science and Technology Information (KISTI) for the support of the computational resources with the Nurion cluster (KSC-2024-CRE-0002, KSC-2025-CRE-0163). A part of the computation was also carried out on the Olaf supercomputer, which is supported by the Institute for Basic Science (IBS) Research Solution Center.

This work makes use of the following software: \texttt{PYTHON} \citep{van2009python}, \texttt{NUMPY} \citep{2020Natur.585..357H}, \texttt{SCIPY} \citep{2020NaMet..17..261V}, \texttt{ASTROPY} \citep{2013A&A...558A..33A,2018AJ....156..123A}, \texttt{MATPLOTLIB} \citep{Hunter2007}, and \texttt{JUPYTER} \citep{kluyver2016jupyter}

\end{acknowledgements}

\section*{Data Availability}
Data directly related to this publication are available on reasonable request from the corresponding author. The TNGC-SIDM simulation suite will be made publicly available in a future release, following the data release policy of the TNG-Cluster project. 

\bibliographystyle{aa}
\bibliography{reference}

\begin{appendix}
\section{Relic and halo separation during the minor merger}
\label{app:minor}

\label{sec:x-axis_projection}
\begin{figure}
 \centering
 \includegraphics[width=0.9\columnwidth]{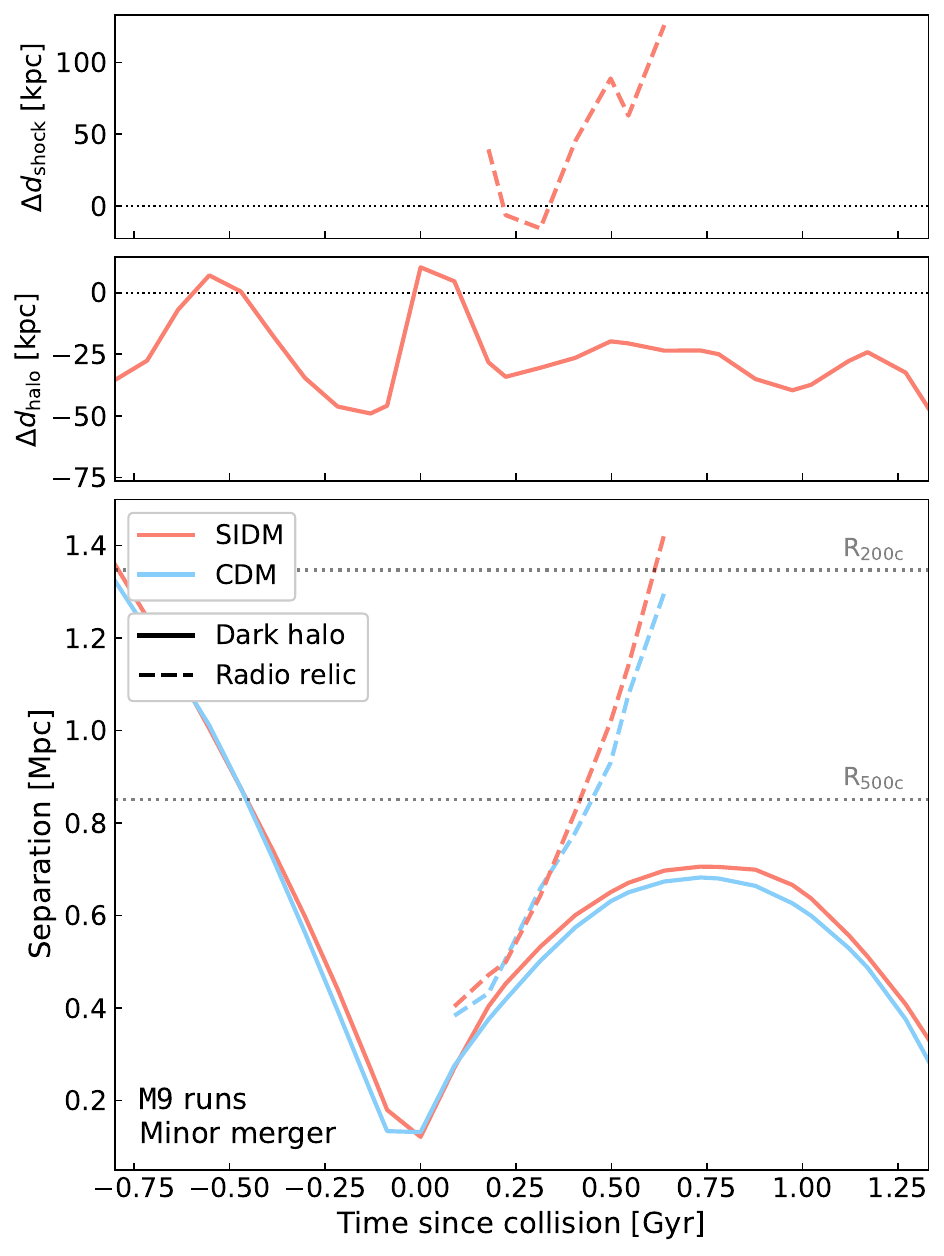}
 \caption{Same as Figure~\ref{fig:DH-relic}, but for the minor merger in the \texttt{M9} runs. As a single radio relic appears in front of the sub-cluster during the minor merger, the relic separation is defined as the separation between the edge of the radio relic and the main cluster center. The halo separation is larger in the SIDM run, and the relic separation is also affected by this difference, as it is measured from the main cluster center.}
 \label{fig:DH-relic-minor}
\end{figure}

Figure~\ref{fig:DH-relic-minor} presents the time evolution of the halo-to-halo separation and relic-to-halo separation measured in the minor merger of the massive cluster simulations (\texttt{M9} runs). As minor mergers only present a single prominent radio relic in front of the sub-cluster motion, we define the relic separation using the separation between the relic edge and the main cluster center, which is known to present time evolution like the double radio relic separations with a larger uncertainty \citep[][]{2026ApJ...998..201L}.

The time evolution of both separations differs from that of the major merger. Consistent with the halo separation degeneracy already identified for minor mergers in \textsection\ref{sec: SIDM_Merger}, the halo separation is larger in the \texttt{M9S1} run than in the \texttt{M9S0} run, presenting a $\Delta d_{\rm halo}$ opposite in sign to that of the major merger, with a $\mytilde20~\rm kpc$ difference across the merger history. We note that the exact trend in the halo separation varies from what is discussed in Figure~\ref{fig:Halo_orbits}, as the separation in Figure~\ref{fig:DH-relic-minor} is measured in the projection map. For the relic separation, the single relic cannot provide a reliable reference for SIDM, since the measurement is anchored to the main cluster center rather than a second shock front and therefore inherits the halo separation difference from the self-interaction. As a result, although the relic separation still evolves linearly with time, its slope is consistently higher in the \texttt{M9S1} run than in the \texttt{M9S0} run, making it an unreliable reference to test self-interactions.

This result highlights that constraining $\sics$ with radio relic separation requires selecting major mergers with double relics, rather than minor merger systems. Although the difference in the single relic separation is small, it grows with time, so averaging over a large sample would not mitigate this. Moreover, even when the TSC is accurately constrained independently of SIDM through the presence of double radio relics, the halo separation in a minor merger presents only a small difference between dark matter models. We therefore conclude that both the halo separation and the single relic separation are unreliable probes of $\sics$ in the minor merger, and that testing the self-interacting nature of dark matter with radio relics specifically requires major cluster mergers.

\section{Spatial offset between BCG and total mass distribution}
\label{app:BCGTotal}

We discuss the spatial offset between the collisionless BCG and the self-interacting dark matter halo in \textsection\ref{sec:offset}. In observations, however, the dark matter distribution is probed with gravitational lensing, which traces the total projected mass distribution instead of the dark matter mass map. As shown in Figure~\ref{fig:SIDMhalo}, the stellar density exceeds the dark matter density in the central region of the SIDM cluster due to its isothermal core, and so the centroid of the projected mass distribution is expected to follow the stellar mass distribution. We therefore test how the measured spatial offset changes when the total mass distribution is used.

\begin{figure}
 \centering
 \includegraphics[width=0.9\columnwidth]{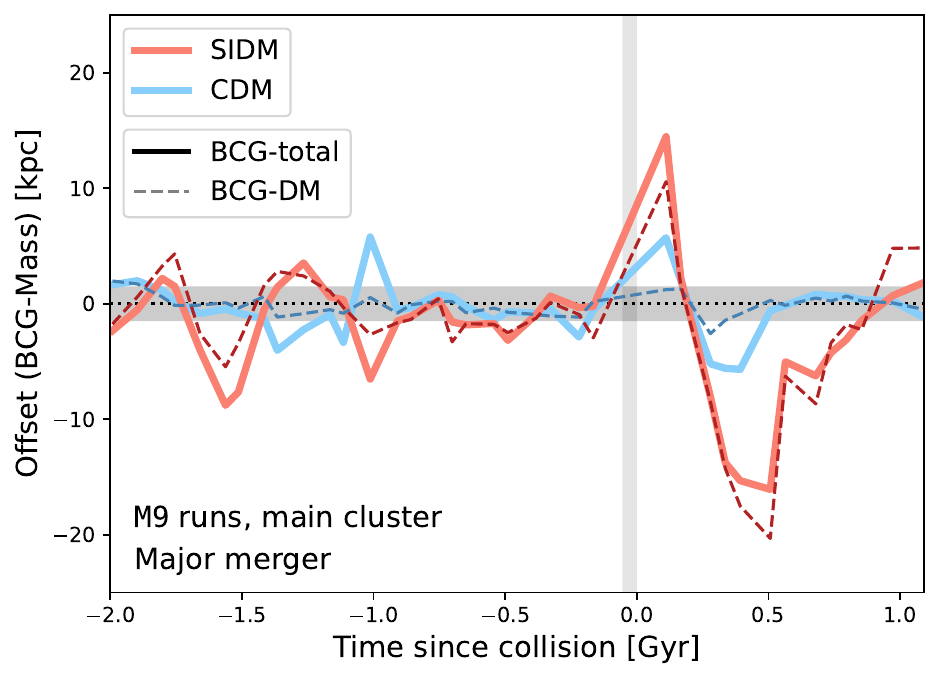}
 \caption{Same as the top panel of Figure~\ref{fig:DMBCGoffset}, but using the total projected mass map instead of the dark matter mass distribution. The thick lines show the spatial offset between the BCG and the centroid of the projected dark matter, gas, and stellar mass smoothed with a 50 kpc Gaussian kernel, measured in the main cluster of the M9 major merger for the SIDM with $\sics=1\rm~cm^2~g^{-1}$ (red) and CDM (blue) runs. The thin dashed lines show the BCG-dark matter halo offset without smoothing from Figure~\ref{fig:DMBCGoffset} for comparison. The BCG-total mass offset follows the oscillation of the BCG-dark matter halo offset in the SIDM run, while the CDM run shows a larger offset than in the unsmoothed measurement.}
 \label{fig:BCGoffset_total}
\end{figure}

In Figure~\ref{fig:BCGoffset_total}, we repeat the measurement in Figure~\ref{fig:DMBCGoffset}, but define the spatial offset between the BCG and the center of the total mass distribution, which includes dark matter, gas, and stars. The total mass distribution has been smoothed with a Gaussian kernel with a width of 50~kpc to mimic the resolution of weak-lensing mass reconstruction. The offset follows the evolutionary trend shown in Figure~\ref{fig:DMBCGoffset}, and the BCG of the main cluster shows a similar oscillatory behavior with time. The main cluster reaches an offset of $\mytilde20\rm~kpc$ at $\rm TSC\sim0.5\rm~Gyr$, similar to the offset measured from the dark matter distribution. This result suggests that the spatial offset would remain visible in gravitational lensing of the total mass distribution.

The smoothed total mass map can recover the oscillatory behavior of the BCG, as dark matter dominates the mass distribution on the smoothing scale. As shown in Figure~\ref{fig:SIDMhalo}, the stellar mass density is larger than the dark matter density within $\lesssim15\rm~kpc$, and beyond this scale, dark matter dominates the mass distribution. As a result, the center of mass defined with the unsmoothed total mass distribution is located closer to the BCG, while the distribution smoothed with a large kernel follows the dark matter and so recovers the BCG-dark matter halo offset. We note that the main cluster in the CDM run shows a larger offset with the BCG in the total mass distribution, as the extended mass distribution in the smoothed map shifts the centroid away from the BCG.

\begin{figure}
 \centering
 \includegraphics[width=0.9\columnwidth]{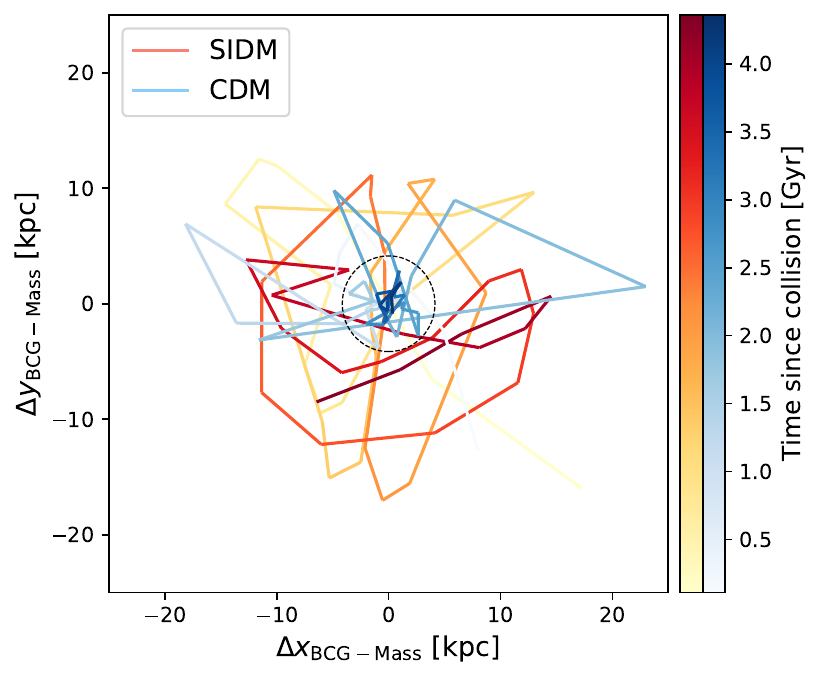}
 \caption{Same as Figure~\ref{fig:Longer_offset}, but using the total projected mass map of dark matter, gas, and stars smoothed with a 50 kpc Gaussian kernel instead of the dark matter mass distribution. The red and blue lines denote the SIDM with $\sics=1\rm~cm^2~g^{-1}$ and CDM runs, respectively. The offset in the SIDM run is smaller than in Figure~\ref{fig:Longer_offset}. Still, the spatial offset in the SIDM run is larger than that of the CDM run when the merger is completed.}
 \label{fig:longer_offset_total}
\end{figure}

Figure~\ref{fig:longer_offset_total} shows the long-term spatial offset between the BCG and the center of the total mass distribution. Compared with Figure~\ref{fig:Longer_offset}, the offset in the SIDM run is smaller, while the offset in the CDM run becomes larger and occasionally exceeds that of the SIDM run. As we describe in \textsection\ref{sec:offset}, the large offset appears when the main cluster experiences the pericenter passage with the sub-cluster. Thus, smoothing shifts the centroid of the total mass towards the sub-cluster, while the stellar centroid is less affected, as the stars are more concentrated than the dark matter (Figure~\ref{fig:DMstar_dist}). At the late merger phase with $\rm TSC>3\rm~Gyr$, when the merger is complete, the offset in the CDM run is smaller than the gravitational softening length, whereas the SIDM run still shows an offset of $\mytilde10\rm~kpc$.

This result indicates that the mass centroid identified from the smoothed total mass distribution, as in gravitational lensing, can still detect the spatial offset discussed in this work. Nevertheless, it also highlights the potential contribution of interlopers. As shown in Figure~\ref{fig:longer_offset_total}, the BCG-total mass offset in the CDM run is large when the sub-cluster passes close to the main cluster center. As observations cannot exclude interloping mass as we do in our analysis, careful target selection that avoids interloping mass structures would be required.

\end{appendix}
\end{document}